\documentclass[floats,floatfix,twocolumn,showpacs,amssymb,prd,superscriptaddress,nofootinbib]{revtex4-1}
\usepackage{graphicx, epsfig, amssymb} 
\usepackage{amsmath, amsfonts}
\usepackage{bm} 
\usepackage[linktocpage]{hyperref}
\usepackage[caption=false]{subfig}
\usepackage[usenames]{color}

\def\be{\begin{equation}}
\def\ee{\end{equation}}
\def\beq{\begin{eqnarray}}
\def\eeq{\end{eqnarray}}

\newcommand{\bea}{\begin{eqnarray}}
\newcommand{\eea}{\end{eqnarray}}
\newcommand{\ben}{\begin{enumerate}}
\newcommand{\een}{\end{enumerate}}
\newcommand{\bi}{\begin{itemize}}
\newcommand{\ei}{\end{itemize}}

\newcommand{\nn}{\nonumber}

\begin{document}

\title{\large Compact stars in alternative theories of gravity.\\
  Einstein-Dilaton-Gauss-Bonnet gravity}

\author{Paolo Pani}\email{paolo.pani@ist.utl.pt}
\affiliation{CENTRA, Departamento de F\'{\i}sica, 
Instituto Superior T\'ecnico, Universidade T\'ecnica de Lisboa - UTL,
Av.~Rovisco Pais 1, 1049 Lisboa, Portugal.}

\author{Emanuele Berti} 
\affiliation{Department of Physics and Astronomy, The University of
  Mississippi, University, MS 38677, USA.}  
\affiliation{California Institute of Technology, Pasadena, CA 91109,
  USA}

\author{Vitor Cardoso} 
\affiliation{CENTRA, Departamento de F\'{\i}sica, 
Instituto Superior T\'ecnico, Universidade T\'ecnica de Lisboa - UTL,
Av.~Rovisco Pais 1, 1049 Lisboa, Portugal.}
\affiliation{Department of Physics and Astronomy, The University of
  Mississippi, University, MS 38677, USA.}

\author{Jocelyn Read} 
\affiliation{Department of Physics and Astronomy, The University of
  Mississippi, University, MS 38677, USA.}

\date{\today} 

\begin{abstract} 
We develop a theoretical framework to study slowly rotating compact
stars in a rather general class of alternative theories of gravity,
with the ultimate goal of investigating constraints on alternative
theories from electromagnetic and gravitational-wave observations of
compact stars. Our Lagrangian includes as special cases scalar-tensor
theories (and indirectly $f(R)$ theories) as well as models with a
scalar field coupled to quadratic curvature invariants. As a first
application of the formalism, we discuss (for the first time in the
literature) compact stars in Einstein-Dilaton-Gauss-Bonnet gravity. We
show that compact objects with central densities typical of neutron
stars cannot exist for certain values of the coupling constants of the
theory. In fact, the existence and stability of compact stars sets
more stringent constraints on the theory than the existence of black
hole solutions. This work is a first step in a program to
systematically rule out (possibly using Bayesian model selection)
theories that are incompatible with astrophysical observations of
compact stars.
\end{abstract}


\maketitle
\date{today}

\section{Introduction}

\noindent{\bf{\em Compact stars as nuclear physics laboratories.}}
Studies of compact stars in general relativity have been textbook
material for decades \cite{Misner:1974qy,Shapiro:1983du}.
Neutron stars can be considered ``cold'' by nuclear physics standards,
so their mass-radius relation $M(R)$ is uniquely determined by the
equation of state (EOS) of matter at high densities, i.e. by the
relation between pressure and energy density $P(\rho)$. From an
observational point of view, one usually assumes general relativity to
be correct. Under this assumption (which of course is backed up by a
wealth of observational evidence \cite{Will:2005va}), the Holy Grail
of astronomical observations of neutron stars is the determination of
the EOS from measurements of macroscopic properties such as masses,
radii and moments of inertia.

Better observational estimates of neutron star masses and radii are
progressively improving our understanding of the EOS.  Lindblom
\cite{1992ApJ...398..569L} presented a concrete scheme for
reconstructing $P(\rho)$ from observations of $M(R)$. More recently,
Read {\it et al.}  \cite{Read:2008iy} approximated the high-density
EOS by piecewise polytropic models, showing that current astrophysical
measurements yield stringent constraints on the piecewise polytropic
parameters. In the same spirit, Lindblom \cite{Lindblom:2010bb}
proposed to replace piecewise polytropes by spectral expansions, that
should give a more faithful representation of the EOS.

Our understanding of the functional form of the EOS from observed
masses and radii has made impressive strides in the recent past
\cite{Ozel:2009da,Ozel:2010fw,Steiner:2010fz} (see also
\cite{Lattimer:2006xb,Lattimer:2010zz} for reviews).  Demorest {\it et
  al.}  \cite{Demorest:2010bx} recently determined a value $M=1.97\pm
0.04~M_\odot$ for the mass of PSR J1614-2230, a pulsar in a white
dwarf-neutron star binary system.
This precisely measured mass is large enough to rule out many
candidate EOSs \cite{2010ApJ...724L.199O,Lattimer:2010uk}. Vice versa,
theoretical progress in microscopic calculations based on chiral
effective field theory is leading to a better understanding of
neutron-rich matter below nuclear densities, and hence to more
stringent constraints on the mass-radius relationship
\cite{Hebeler:2010jx}.

\noindent{\bf{\em Compact stars as strong gravity laboratories.}}
Most studies of the possibility of reconstructing the EOS from compact
star observations assume that general relativity is the correct theory
of gravity.  General relativity passed all observation tests so far
\cite{Will:2005va}, but the ``real'' theory of gravity may well differ
significantly from it in strong field regions. In fact, cosmological
observations and conceptual difficulties in quantizing Einstein's
theory suggest that general relativity may require modifications.

Compact stars are an ideal natural laboratory to look for possible
modifications of Einstein's theory and their observational signatures
\cite{Psaltis:2008bb}. Besides ruling out specific models for the EOS,
experiments may (and should) try to rule out also alternative theories
of gravity that are unable to explain observations.  A comprehensive
study of how EOS models and alternative theories affect macroscopic
observable quantities of compact stars requires a Bayesian model
selection framework, where one compares the predictions of any
specific theory of gravity (and of different EOS models) against the
growing body of observational data. Of course, an important
prerequisite of any such analysis is the construction of stellar
models in the largest possible family of alternative theories of
gravity that are not ruled out by weak-field experiments, cosmological
constraints or observations of compact binary systems. The present
work is a first step in this direction.

The plan of the paper is as follows. In Sec.~\ref{sec:altstars}, to
put our work in context, we briefly review some studies of compact
stars in alternative theories of gravity. In Sec.~\ref{sec:EST} we
present the Lagrangian for what we call ``extended scalar-tensor
theories''. In Sec.~\ref{sec:starsEST} we write down the equations
describing the structure of static and slowly-rotating stars in this
class of theories, and we discuss our chosen models for the EOS. In
Sec.~\ref{sec:starsEDGB} we present numerical results and discuss
their implications. We conclude by discussing extensions of the
present work to other theories and comparisons with observations. 
Unless stated otherwise, we use geometrical units ($G=c=1$).

\section{Stars in alternative theories: a brief review\label{sec:altstars}}

The study of compact stars in alternative theories of gravity has a
long history. In this paper we begin a systematic exploration of
stellar structure in a large class of modified gravity theories. Far
from providing a comprehensive review, here we point to some relevant
literature, mainly to put our work in context.

\noindent{\bf{\em i) Scalar-tensor theories.}}
Scalar-tensor gravity is one of the simplest and best-motivated
modifications of general relativity, because scalar fields are
predicted by almost all attempts to incorporate gravity into the
standard model \cite{Fujii:2003pa}. Therefore it should come as no
surprise that most work on stellar structure concerns variants of
scalar-tensor theory. The equations of hydrostatic equilibrium in the
best-known variant of scalar-tensor theories (Brans-Dicke theory) were
first studied by Salmona \cite{Salmona:1967zz}. Soon after, Nutku
\cite{1969ApJ...155..999N} explored the radial stability of stellar
models using a post-Newtonian treatment.  Hillebrandt and Heintzmann
\cite{1974GReGr...5..663H} analyzed incompressible (constant density)
configurations. Zaglauer \cite{Zaglauer:1992bp} carried out a detailed
calculation of the so-called ``sensitivities'' of neutron stars, which
determine the amount of dipolar gravitational radiation emitted by
compact binaries in scalar-tensor theories \cite{Will:1989sk}. Most of
these studies found that corrections to neutron star structure are
suppressed by a factor $1/\omega_{\rm BD}$, where $\omega_{\rm BD}$ is
the Brans-Dicke coupling constant. At present, the most stringent
bound on this parameter ($\omega_{\rm BD}>40,000$) comes from Cassini
measurements of the Shapiro time delay \cite{Will:2005va}.

As pointed out by Damour and Esposito-Far\'ese \cite{Damour:1993hw},
the coupling of the scalar with matter can produce a ``spontaneous
scalarization'' phenomenon by which certain ``generalized''
scalar-tensor theories may pass all weak-field tests, and at the same
time introduce macroscopically (and observationally) significant
modifications to the structure of compact stars. More detailed studies
of stellar structure \cite{Damour:1996ke,Salgado:1998sg}, numerical
simulations of collapse
\cite{Shibata:1994qd,Harada:1996wt,Novak:1997hw} and a stability
analysis \cite{Harada:1997mr} confirmed that ``spontaneously
scalarized'' configurations would indeed be the end-state of stellar
collapse in these theories. In fact, spontaneously scalarized
configurations may arise as a result of semiclassical vacuum
instabilities \cite{Pani:2010vc}. Tsuchida {\it et al.}
\cite{Tsuchida:1998jw} extended the Buchdahl inequality ($M/R\leq 4/9$
for incompressible stars) to generalized scalar-tensor theories. For a
comprehensive study of analytic solutions and an extensive
bibliography, see \cite{Horbatsch:2010hj}.

\noindent{\bf{\em ii) f(R) theories.}}
Theories that replace the Ricci scalar $R$ by a generic function
$f(R)$ in the Einstein-Hilbert action\footnote{Here and below we refer
  to $f(R)$ theories in the metric formalism. Theories of the Palatini
  type have conceptual problems: for example, spherically symmetric
  polytropic ``stars'' present curvature singularities
  \cite{Barausse:2007pn,Barausse:2007ys}.} can always, at least in
principle, be mapped into scalar-tensor theories
\cite{DeFelice:2010aj}.  The existence of compact stars in metric
$f(R)$ models that have been proposed to explain cosmological
observations, such as the Starobinsky model \cite{Starobinsky:2007hu},
was studied by many authors
\cite{Kobayashi:2008tq,Upadhye:2009kt,Babichev:2009td,
  Babichev:2009fi,Miranda:2009rs} with controversial results. One
possible explanation of the partial disagreement between different
authors is that the mapping between $f(R)$ theories and scalar-tensor
theories is in general multivalued, and therefore one should be
careful when considering the scalar-tensor ``equivalent'' of an $f(R)$
theory \cite{Jaime:2010kn}. A perturbative approach to stellar
structure in $f(R)$ gravity is also possible \cite{Cooney:2009rr}.

\noindent{\bf{\em iii) Higher-curvature gravity.}}
Besides theories where the Lagrangian is a generic function of $R$, it
is of interest to consider theories where the Lagrangian is built out
of quadratic \cite{Yunes:2011we} (or even higher-order) contractions
of the Riemann and Ricci tensors.  As we explain below, the
requirement that the field equations should be second-order means that
quadratic corrections must appear in the Gauss-Bonnet (GB) combination
\begin{equation}
{\cal R}^2_\text{GB}
=R^2-4R_{ab}R^{ab}+R_{abcd}R^{abcd}\,,\label{GBterm} 
\end{equation}
where $R_{abcd}$ is the Riemann tensor and $R_{ab}$ is the Ricci
tensor. Since the GB term in four dimensions is a topological
invariant, the GB combination introduces modifications to general
relativity only when coupled to a nonzero scalar field or other forms
of matter. The simplest and better motivated case\footnote{In analogy
  with $f(R)$ models, $f({\cal R}^2_\text{GB})$ models have been
  studied in a cosmological context. Observational constraints on
  $f({\cal R}^2_\text{GB})$ models are quite tight (see e.g. Sections
  12.3 and 12.4 of \cite{DeFelice:2010aj}) and we will not include
  them in our analysis.} is Einstein-Dilaton-Gauss-Bonnet (EDGB)
gravity \cite{Kanti:1995vq}, where the GB term is coupled to a
dynamical scalar field, the dilaton.  The EDGB correction to the
Einstein-Hilbert action appears in low-energy, tree-level effective
string theory \cite{Gross:1986mw}.

The study of EDGB gravity in relativistic astrophysics has been
limited to a mathematical analysis of black hole solutions
\cite{Mignemi:1992nt,Kanti:1995vq,Torii:1996yi,Alexeev:1996vs} and,
more recently, to their possible observational signatures
\cite{Pani:2009wy,Yunes:2011we,Kleihaus:2011tg}.  To our knowledge,
the present study is the first investigation of compact stars in the
theory.  Static black holes in EDGB gravity only exist when $\alpha>0$
\cite{Kanti:1995vq}. Hence, we shall restrict our study to the case of
positive $\alpha$.

\noindent{\bf{\em iv) Parity-violating theories.}} Chern-Simons gravity is
the simplest theory that allows for parity-violating corrections to
general relativity \cite{Alexander:2009tp}. Due to the nature of the
Chern-Simons corrections, all spherically symmetric solutions of
Einstein's theory are also solutions of Chern-Simons gravity. However,
spinning objects in the nondynamical \cite{Smith:2007jm} and dynamical
\cite{Yunes:2009ch} versions of the theory are affected by the
Chern-Simons coupling. Future observations of the moment of inertia of
compact stars may strongly constrain the parameters of the theory
\cite{Yunes:2009ch}.

\noindent{\bf{\em v) Lorentz-violating theories.}} Einstein-aether
theory introduces a dynamical unit timelike vector coupled to gravity
as a natural way to implement Lorentz violation in Einstein's
theory. In the parameter space compatible with Solar System
constraints, spherically symmetric neutron stars in Einstein-aether
theory have a lower maximum mass than in general relativity
\cite{Eling:2006df,Eling:2007xh}. Another popular Lorentz-violating
theory is Ho$\check{\rm r}$ava gravity.  The matching conditions
necessary to obtain stellar solutions in this theory were considered
in \cite{Greenwald:2009kp}, but (to our knowledge) there are no
phenomenological studies of compact stars using realistic EOS models.

\noindent{\bf{\em vi) Massive gravity.}} Recently, Damour {\it et al.}
\cite{Damour:2002gp} reconsidered the discontinuity problem of massive
gravity and its possible resolution through Vainshtein's nonlinear
resummation of nonlinear effects. As part of this study, the authors
investigated the viability of spherically symmetric stars in the
theory. They showed that some solutions show physical singularities,
but also that there exist regular solutions interpolating between a
modified general relativistic interior and a de Sitter exterior, with
curvature proportional to the square of the putative graviton mass. A
more phenomenological study of observational constraints (including
stellar rotation) is still lacking.

\noindent{\bf{\em vii) Eddington inspired gravity.}} Ba\~nados and Ferreira
\cite{Banados:2010ix} recently proposed a theory that is equivalent to
general relativity in vacuum, but differs from it in the coupling with
matter. An interesting aspect of this theory is that singularities
cannot form in early cosmology and during gravitational collapse
\cite{Banados:2010ix,Pani:2011mg}. The maximum mass of compact stars
in the observationally viable sector of Eddington-inspired gravity may
be larger than in general relativity, even for ``ordinary'' EOS models
\cite{Pani:2011mg}.

\noindent{\bf{\em viii) Gravitational aether, f(T), TeVeS and other
    theories.}}
Some alternatives to general relativity that were proposed to explain
cosmological observations have also been analyzed, at least to some
extent, in the context of compact stars.  Among these theories we can
list ``gravitational-aether'' theory \cite{Kamiab:2011am}, $f(T)$
gravity \cite{Boehmer:2011gw} and Bekenstein's TeVeS
\cite{Bekenstein:2004ne}.
In higher-dimensional braneworld models, the embedding of
four-dimensional stellar solutions ``on the brane'' within acceptable
higher-dimensional solution is a nontrivial problem
\cite{Maartens:2010ar} (but see \cite{Wiseman:2001xt} for related work
in a slightly different context).

\section{Extended scalar-tensor theories\label{sec:EST}}

From the previous summary it should be clear that it is nearly
impossible to discuss {\em all} strong-field modifications of general
relativity in a unified framework. However, in this section we show
that on the basis of some rather general arguments we can easily write
down a Lagrangian encompassing the first four classes of theories
reviewed above (namely general scalar-tensor theories, $f(R)$
theories, EDGB gravity and Chern-Simons gravity).

Our starting point is a Lagrangian in which gravity is coupled to a
single (generically charged) scalar field $\phi$ in all possible ways,
including all linearly independent quadratic curvature corrections to
general relativity. We call these models ``extended scalar-tensor
theories''.  The most general Lagrangian of such a theory contains
several functions of the scalar field in the combination
\begin{eqnarray}
{\cal L}&&= f_0(|\phi|)R-\gamma(|\phi|)\partial_a\phi^*\partial^a\phi-V(|\phi|)\nn\\
&&+ f_1(|\phi|)R^2+ f_2(|\phi|)R_{ab}R^{ab}+ f_3(|\phi|)R_{abcd}R^{abcd}\nn\\
&&+ f_4(|\phi|)R_{abcd}{}^*R^{abcd}+{\cal
L}_\text{mat}\left[\Psi,A^2(|\phi|)g_{ab}\right]\,,
\label{lagrangian}
\end{eqnarray}
where ${}^*R^{abcd}$ is the dual of the Riemann tensor, which
introduces possible parity-violating corrections
\cite{Alexander:2009tp}.
From the Lagrangian above, the equations of motion read:
\begin{equation}
G_{ab}+\frac{1}{f_0}\left[{\cal H}_{ab}+{\cal I}_{ab}+{\cal J}_{ab}+{\cal K}_{ab}\right]=\frac{1}{2
f_0}\left[A^2T_{ab}^\text{mat}+T_{ab}^{(\phi)}\right]\,,
\end{equation}
where $T_{ab}^\text{mat}={2}{(-g)^{-1/2}}\delta S_m/\delta g_{ab}$ is
the matter stress-energy tensor in the Jordan frame,
\begin{eqnarray}
T_{ab}^{(\phi)}&&=\gamma\left[2\partial_{(a}\phi^*\partial_{b)}\phi-g_{ab}
\partial_c\phi^*\partial^c\phi\right]-g_{ab}V\nn\\
&&+2\nabla_a\nabla_b f_0-2g_{ab}\nabla^2 f_0  \,,
\end{eqnarray}
and, following the notation of Ref. \cite{Yunes:2011we}, we have
defined
\begin{widetext}
\begin{subequations}
\begin{eqnarray}
{\cal{H}}_{ab} &\equiv& -4
v_{(a}^{(1)} \nabla_{b)} R - 2 R \nabla_{(a} v_{b)}^{(1)} + g_{ab} \left[2
R\nabla^{c} v_{c}^{(1)} + 4 v_{c}^{(1)} \nabla^{c}R \right]+ f_1 \left[ 2 R_{ab} R - 2\nabla_{ab} R
- \frac{1}{2}
  g_{ab} \left[ R^2 - 4\square R \right] \right] \,,\nn
\\
{\cal{I}}_{ab} &\equiv& -v_{(a}^{(2)} \nabla_{b)} R - 2v_c^{(2)} \left[
  \nabla_{(a} R_{b)}{}^c - \nabla^c R_{ab} \right] +\nabla^c v_c^{(2)}R_{ab}- 2 R_{c(a}\nabla^c
v_{b)}^{(2)} + g_{ab} \left[ v_c^{(2)}
  \nabla^c R + R^{cd} \nabla_c v_d^{(2)} \right] \nn\\
&+& f_2 \left[ 2 R^{cd} R_{acbd} - \nabla_{ab} R + \square R_{ab}
+ \textstyle{\frac{1}{2}}g_{ab} \left[  \square R - R_{cd} R^{cd}
  \right] \right] \,,\nn
  \\
{\cal{J}}_{ab}
&\equiv& - 8 v_c^{(3)} \left[ \nabla_{(a} R_{b)}{}^c - \nabla^{c} R_{ab}\right] + 4
R_{acb}^d \nabla^{c} v_{d}^{(3)} - f_3 \left[ 2 \left[ R_{ab} R - 4R^{cd} R_{acbd} + \nabla_{ab}
  R - 2 \square R_{ab} \right]-\textstyle{\frac{1}{2}} g_{ab} \left[ R^2 - 4 R_{cd}
    R^{cd}\right] \right] \,, \nn\\
{\cal{K}}_{ab}
&\equiv& 4 v_{c}^{(4)} \epsilon^c{}_{de(a} \nabla^{e} R_{b)}{}^d + 4
\nabla_{d}v_{c}^{(4)} {}^{*}R_{(a}{}^c{}_{b)}{}^d\,,\nn
\end{eqnarray}
\end{subequations}
\end{widetext}
with $v_{a}^{(i)} \equiv \nabla_{a} f_i(|\phi|)$,
$\nabla_{ab}=\nabla_a\nabla_b$, $\square=\nabla_a\nabla^a$ and
$\epsilon^{abcd}$ the Levi-Civita tensor.
\begin{table*}[htb]
\caption{\label{tab:particular} Specific models obtained from the
  Lagrangian~\eqref{lagrangian}. Here $\kappa\equiv (16\pi G)^{-1}$.}

\begin{tabular}{ccccccccccc}
\hline
\hline
& $f_0$ & $f_1$ & $f_2$ & $f_3$ & $f_4$ & $\omega$ & $V$ & $\gamma$ & $A$ & ${\cal L}_\text{mat}$ \\
\hline
General relativity& $\kappa$			& 0 & 0 & 0 & 0 & 0		& 0	&1&1	
	& perfect fluid\\
Scalar-tensor (Jordan frame) \cite{Damour:1996ke}	& $F(\phi)$ 	& 0 & 0 & 0 & 0 & 0 	
& $V(\phi)$ 		&$\gamma(\phi)$&1		& perfect fluid \\ 
Scalar-tensor (Einstein frame) \cite{Damour:1993hw}	& $\kappa$ 	& 0 & 0 & 0 & 0 & 0 	
& $V(\phi)$ 		&$2\kappa$&$A(\phi)$		& perfect fluid \\ 
$f(R)$ \cite{DeFelice:2010aj}		& $\kappa$		& 0 & 0 & 0 & 0	& 0 & $\kappa\frac{R
f_{,R}-f}{16\pi \bar G f_{,R}^2}$ &$2\kappa$&$f_0^{-1/2}=f_{,R}^{-1/2}$ & perfect fluid\\
Quadratic gravity \cite{Yunes:2011we} 		& $\kappa$ 			& $\alpha_1\phi$ &
$\alpha_2\phi$ & $\alpha_3\phi$ & $\alpha_4\phi$ & 0 		& $0$ 	&1&1			&
perfect fluid \\ 
EDGB \cite{Kanti:1995vq} 		& $\kappa$ 			& $e^{\beta\phi}$ & $-4f_1$
& $f_1$ & 0 & 0 		& $0$ 	&1&1			& perfect fluid \\ 
Dynamical Chern-Simons \cite{Yunes:2009ch}	 	& $\kappa$ 			& 0 & 0 & 0
& $\beta\phi$ & 0 		& $0$ 	&1&1			& perfect fluid \\ 
Boson stars \cite{Ruffini:1969qy} 	& $\kappa$ 			& 0 & 0 & 0 & 0 & $\omega$ 
& $\frac{m^2}{2}|\phi|^2$ &1&1	& $0$ \\
\hline
\hline
\end{tabular}
\end{table*}
The modified Klein-Gordon equation reads
\begin{eqnarray}
\square\phi&&=\frac{\phi}{2|\phi|\gamma}\left[
V'-\gamma'\partial_a\phi^*\partial^a\phi-f_0'R-f_1'R^2-f_2'R_{ab}R^{ab}\right.\nn\\
&&\left.-f_3'R_{abcd}R^{abcd}-f_4'R_{abcd}{}^* R^{abcd}
-{A'}A^3T^\text{mat}\right]\,,\label{KG}
\end{eqnarray}
together with its complex conjugate. In the equations above, a prime
denotes a derivative with respect to $|\phi|$.

As shown in Table~\ref{tab:particular}, this theory is sufficient to
discuss stellar structure in many of the alternative theories listed
in Sec.~\ref{sec:altstars} (and it can also describe boson stars in
general relativity, if we work in vacuum).  As a matter of fact, some
terms in the Lagrangian \eqref{lagrangian} are redundant. For example,
in ordinary scalar-tensor theories the functions $f_0$ and $\gamma$
can be removed via a conformal transformations of the metric and a
redefinition of the scalar field; i.e., by reformulating the theory in
the Einstein frame \cite{Damour:1996ke}. However, depending on the
explicit form of $f_0$ and $\gamma$, these transformations can be hard
(if not impossible) to write in a closed analytic form. For this
reason we find it convenient to start from the general Lagrangian
\eqref{lagrangian}, which reduces to standard scalar-tensor theories
in the Jordan frame if $A(|\phi|)\equiv1$ and in the Einstein frame if
$f_0(|\phi|)\equiv1/(16\pi)$ and $\gamma(|\phi|)\equiv1$.  Another
advantage of this approach is that, at least in principle, it should
also encompass generic $f(R)$ theories, which are equivalent to
particular scalar-tensor theories (but see \cite{Jaime:2010kn} for
possible issues with this point of view).

\subsection{Simplifying the model}
For generic coupling functions, the terms ${\cal{H}}_{ab}$,
${\cal{I}}_{ab}$, ${\cal{J}}_{ab}$ appearing on the left-hand side of
the equations of motion introduce higher-order derivatives of the
metric functions, unless quadratic terms in the curvature enter the
action in the GB combination~\eqref{GBterm}. The GB combination
corresponds to setting $f_2=-4f_1$ and $f_3=f_1$ in our model. Thus,
if we only want second-order equations of motion the
Lagrangian~\eqref{lagrangian} must reduce to
\begin{eqnarray}
{\cal L}&&= f_0(|\phi|)R+ f_1(|\phi|){\cal R}^2_\text{GB}+ f_4(|\phi|)R_{abcd}{}^*R^{abcd}\nn\\
&&-\gamma(|\phi|)\partial_a\phi^*\partial^a\phi-V(|\phi|)+{\cal
L}_\text{mat}\left[\Psi,A^2(|\phi|)g_{ab}\right]\,.\label{lagrangianGB}
\end{eqnarray}
In order to avoid the complications related to higher-order
derivatives, from now on we will specialize to this Lagrangian.

\section{Perfect fluid compact stars in extended scalar-tensor theories\label{sec:starsEST}}

\subsection{Static solutions}

We begin by looking for static, spherically symmetric equilibrium
solutions of the field equations with metric
\begin{eqnarray}
ds^2_0=-B(r)dt^2+\frac{dr^2}{1-2m(r)/r}+r^2d\theta^2+r^2\sin^2\theta\,d\varphi^2
\nonumber
\end{eqnarray}
and a charged, spherically symmetric scalar field
\be
\phi(t,r)=\Phi(r)e^{-i\omega t}\,. 
\ee
Because of the assumed spherical symmetry, the Pontryagin density
vanishes ($R_{abcd}{}^* R^{abcd}=0$) and the equations of motion do
not depend on $f_4$. Our ansatz for the scalar field also implies that
the Klein-Gordon equation~\eqref{KG} and its conjugate coincide.

We consider perfect-fluid stars with energy density $\rho(r)$ and
pressure $P(r)$ such that
\begin{equation}
 T^{\mu\nu}_{\rm mat}\equiv T^{\mu\nu}_{\rm
perfect\,fluid}=\left(\rho+P\right)u^\mu\,u^\nu+g^{\mu\nu}P\,,\label{Tmunu_fluid}
\end{equation}
where the fluid four-velocity $u^\mu=(1/\sqrt{B},0,0,0)$. Note that
the matter fields are defined in the Jordan frame. The stress-energy
tensor in the Einstein frame reads
$T_{\mu\nu}^{(E)}=A^2(|\phi|)T_{\mu\nu}$.  To close the system, as
usual, we must also specify an EOS $P=P(\rho)$.

The field equations for a static, spherically symmetric perfect-fluid
star in extended scalar-tensor theories read
\begin{widetext}
\begin{eqnarray}
E_{00}&=&-\frac{r^5 \omega ^2 \gamma (\Phi) \Phi ^2}{B}-r^5 V(\Phi)-r^5 A(\Phi)^2 \rho +4 r^3
f_0(\Phi) m'-4 r^4 f_0'(\Phi) \Phi '+6 r^3 m f_0'(\Phi) \Phi '+16 r m f_1'(\Phi) \Phi '\nn\\
&&-48 m^2 f_1'(\Phi) \Phi '+2 r^4 f_0'(\Phi) m' \Phi '-16 r^2 f_1'(\Phi) m' \Phi '+48 r m f_1'(\Phi)
m' \Phi '-r^5 \gamma (\Phi) {\Phi'}^2+2 r^4 m \gamma (\Phi) {\Phi'}^2-2 r^5 {\Phi'}^2
f_0''(\Phi)\nn\\
&&+4 r^4 m {\Phi'}^2 f_0''(\Phi)-16 r^2 m {\Phi'}^2 f_1''(\Phi)+32 r m^2 {\Phi'}^2 f_1''(\Phi)-2 r
(r-2 m) \left[r^3 f_0'(\Phi)+8 m f_1'(\Phi)\right] \Phi ''=0\,,\\
E_{11}&=&r^4 \omega ^2 \gamma (\Phi) \Phi ^2-(r-2 m) B' \left\{2 r^2 f_0(\Phi)+\left[r^3
f_0'(\Phi)-8 (r-3 m) f_1'(\Phi)\right] \Phi '\right\}+r B \left\{4 f_0(\Phi) m+r \left[r^2 A(\Phi)^2
P\right.\right.\nn\\
&&\left.\left.-r^2 V(\Phi)+(r-2 m) \Phi ' \left(-4 f_0'(\Phi)+r \gamma (\Phi) \Phi
'\right)\right]\right\}=0\,,\\
E_\text{cons}&=&4 r^5 A(\Phi) B^2 f_0(\Phi) P A'(\Phi) \Phi '+r^5 A(\Phi)^2 B \left[f_0(\Phi)
\left((P+\rho ) B'+2 B P'\right)-2 B P f_0'(\Phi) \Phi '\right]+\Phi ' \times\nn\\
&&\left\{r f_0(\Phi) (r-2 m) B'^2 \left[r^3 f_0'(\Phi)+8 m f_1'(\Phi)\right]+2 B \left\{r f_0'(\Phi)
\left[-r^4 \omega ^2 \gamma (\Phi) \Phi ^2+(r-2 m) B' \left(r^3
f_0'(\Phi)\right.\right.\right.\right.\nn\\
&&\left.\left.\left.\left.-8 (r-3 m) f_1'(\Phi)\right) \Phi '\right]+f_0(\Phi) \left[-r^3 m B'
f_0'(\Phi)+8 r m B' f_1'(\Phi)-24 m^2 B' f_1'(\Phi)+r^4 B' f_0'(\Phi) m'\right.\right.\right.\nn\\
&&\left.\left.\left.-8 r^2 B' f_1'(\Phi) m'+24 r m B' f_1'(\Phi) m'+r^5 \omega ^2 \Phi ^2 \gamma
'(\Phi)+r^4 \gamma (\Phi) \left(2 r \omega ^2 \Phi +(r-2 m) B' \Phi
'\right)\right.\right.\right.\nn\\
&&\left.\left.\left.-r (r-2 m) \left(r^3 f_0'(\Phi)+8 m f_1'(\Phi)\right) B''\right]\right\}+2 r^2
B^2 \left\{r f_0'(\Phi) \left[r^2 V(\Phi)+(r-2 m) \Phi ' \left(4 f_0'(\Phi)-r \gamma (\Phi) \Phi
'\right)\right]\right.\right.\nn\\
&&\left.\left.+f_0(\Phi) \left[r \left(4 f_0'(\Phi) m'+r \left(-r V'(\Phi)+r \gamma '(\Phi) \Phi
'^2+2 \gamma (\Phi) \left(-\left(-2+m'\right) \Phi '+r \Phi ''\right)\right)\right]-2 m \left(2
f_0'(\Phi)\right.\right.\right.\right.\nn\\
&&\left.\left.\left.\left.+r \left(r \gamma '(\Phi) \Phi '^2+\gamma (\Phi) \left(3 \Phi '+2 r \Phi
''\right)\right)\right)\right)\right]\right\}=0\,,\\
E_\text{scal}&=&r (r-2 m) B'^2 \left[r^3 f_0'(\Phi)+8 m f_1'(\Phi)\right]+2 B \left\{B' \left[r^3
f_0'(\Phi) \left(3 m+r \left(-2+m'\right)\right)+8 (r-3 m) f_1'(\Phi) \left(m-r
m'\right)\right]\right.\nn\\
&&\left.-r^5 \omega ^2 \Phi ^2 \gamma '(\Phi)+r^4 \gamma (\Phi) \left[2 r \omega ^2 \Phi +(r-2 m) B'
\Phi '\right]-r (r-2 m) \left[r^3 f_0'(\Phi)+8 m f_1'(\Phi)\right] B''\right\}\nn\\
&&+2 r^3 B^2 \left\{r^2 A(\Phi)^3 (3 P-\rho ) A'(\Phi)+4 f_0'(\Phi) m'-r^2 V'(\Phi)+\Phi ' \left[-2
\gamma (\Phi) \left(3 m+r \left(-2+m'\right)\right)+r (r-2 m) \gamma '(\Phi) \Phi
'\right]\right.\nn\\
&&\left.+2 r (r-2 m) \gamma (\Phi) \Phi ''\right\}=0\,,
\end{eqnarray}
\end{widetext}
where $E_{00}$ and $E_{11}$ are the $\{0,0\}$ and $\{1,1\}$ components
of the modified Einstein equations, $E_\text{cond}=\nabla_a
T^{a2}\equiv0$ and $E_\text{scal}$ denotes the field equation for
the scalar field.

To construct spherically symmetric and static stellar configurations,
we must solve the system above imposing regularity conditions at the
center of the star, i.e.
\begin{equation}
m(0)=0\,,\quad \rho(0)=\rho_c\,,\quad \Phi(0)=\Phi_c\,,\quad \Phi'(0)=0 \,.
\end{equation}
More in general, any field can be expanded close to the center as
\begin{equation}
 X(r)=X^{(0)}+X^{(1)}r+X^{(2)}r^2+{\cal O}(r^3)\,,\label{series_center}
\end{equation}
where $X$ schematically denotes any of the variables $\rho$, $P$,
$\Phi$, $B$ and $m$.  By using the field equations, all coefficients
$X^{(i)}$ ultimately depend on two parameters only, say
$\rho^{(0)}=\rho_c$ and $\Phi^{(0)}=\Phi_c$. Finally, the value of
$\Phi_c$ is fixed through a shooting method in order to obtain an
asymptotically flat solution\footnote{This condition can be easily
  generalized to allow for nonvanishing values of the scalar field
  ($\Phi\to \Phi_\infty\neq 0$ as $r\to\infty$), like those arising in
  $f(R)$ theories and in other modified gravity theories that want to
  reproduce cosmological dynamics.}: $\Phi\to 0$ as $r\to\infty$.

The outcome of this shooting method is a one-parameter family of
solutions characterized only by the central density $\rho_c$. For any
value of $\rho_c$, we can compute the mass $M$ and the radius $R_s$ of
the star. As usual, the mass is obtained from the asymptotic behavior
at infinity
\begin{equation}
 B(r)\to1-\frac{2M}{r}\,,
\end{equation}
whereas the radius is computed by imposing the usual matching
condition at the stellar surface, $P(R_s)=0$. In the exterior of the
star, $P=\rho=0$.  Finally, the baryonic mass, corresponding to the
energy that the system would have if all baryons were dispersed to
infinity, is defined as
\be 
\bar{m}=m_b\int d^3x\sqrt{-g}u^0 n(r)\,, 
\ee
where $n$ denotes the baryonic number.  The normalized binding energy
$\bar{m}/M-1$ is positive for bound (but not necessarily stable)
configurations.

\subsection{Slowly rotating models}\label{sec:slowlyrot}
Once a static stellar model is known, it is easy to construct the
corresponding slowly rotating model by generalizing the classic work
by Hartle \cite{Hartle:1967he}.  For this purpose, we consider the
metric ansatz
\begin{eqnarray}
ds^2&=&-B(r)dt^2+\left[1-2m(r)/r\right]^{-1}dr^2+r^2d\theta^2\nn\\
&+&r^2\sin^2\theta\,\left(d\varphi-\frac{\zeta(r)}{2}dt\right)^2\nn\\
&=&ds^2_0-\zeta(r)r^2\sin^2\theta
dtd\varphi+{\cal O}(\zeta^2)\,.
\label{metric_rot}
\end{eqnarray}
The stress-energy tensor for a rotating fluid can be easily
constructed from Eq.~\eqref{Tmunu_fluid} and the four-velocity
\begin{equation}
 u^\mu=\left\{u^t,0,0,\Omega u^t\right\} \,,\quad 
u^t=\sqrt{-(g_{tt}+2\Omega_{t\varphi}+\Omega^2 g_{\varphi\varphi})}\,,
\label{4velocity_rotating}
\end{equation}
where $\Omega$ is the angular velocity of the fluid.  

Note that, for any nonconstant $f_4(|\phi|)$, the gravitomagnetic part
$g_{t\varphi}$ of the metric would source scalar perturbations through
first order parity-violating terms in Eq.~\eqref{KG}.  If $f_0,f_1,A$
and $\gamma$ are constant and $V\equiv0$, the field equations admit
the same spherically symmetric solutions as in general relativity, and
the background scalar field vanishes.  In this case, the only
first-order corrections arise from the $\{t,\varphi\}$ component of
the Einstein equations and from the (perturbed) scalar equation, since
the stress-energy tensor is quadratic in the scalar field. These two
equations can be solved for $\zeta(r)$ and for the scalar field
perturbation for slow rotation and small scalar fields.  This was done
in Ref.~\cite{Yunes:2009ch}, where the authors studied slowly rotating
neutron stars in dynamical Chern-Simons gravity
($f_4(|\phi|)=\alpha|\phi|$). When (as in our case) the background scalar
field is nonvanishing, all nontrivial Einstein equations would acquire
first-order corrections. Hence, if $f_4(|\phi|)$ is not constant,
Hartle's formalism cannot be applied. In particular, other first-order
metric corrections (in addition to the gravitomagnetic part
$g_{t\varphi}$) must be included in Eq.~\eqref{metric_rot}. This would
result in a system of equations which is very difficult to
decouple. To avoid these issues, from now on we will simply {\em
  assume} that $f_4(|\phi|)=$const.

Using the definition of the four-velocity~\eqref{4velocity_rotating}
and linearizing in the angular velocity $\Omega$, we find that the
solution to the field equations corresponds to the background
(nonrotating) solution plus the solution $\zeta(r)$ of an ordinary
differential equation coming from the $\{t,\varphi\}$ component of the
Einstein equations, namely:
\begin{eqnarray}
\label{eqROT}
&&\zeta''(r)+C_{1}(r)\zeta'(r)=\left[\zeta(r)-\Omega\right]\times\\
&\times&\frac{r^3A(\Phi)^2(P+\rho)}{(r-2M)\left[r^2
    f_0(\Phi)-4(r-2M)f_1'(\Phi)\Phi'\right]}\,,\nn
\end{eqnarray}
where
\begin{eqnarray}
 C_{1}&&=\frac{1}{2 r B (r-2 M) \left(r^2 f_0(\Phi)-4 (r-2 M) f_1'(\Phi) \Phi '\right)}\times\nn\\
&&\left[r (r-2 M) B' \left(4 (r-2 M) f_1'(\Phi) \Phi '-r^2 f_0(\Phi)\right)\right.\nn\\
&&\left.+2 B \left(-r^2 f_0(\Phi) \left(7 M+r \left(M'-4\right)\right)+(r-2
M)\times\right.\right.\nn\\
&&+\left.\left.\times\left(\Phi ' \left(r^3 f_0'(\Phi)+12 f_1'(\Phi) \left(M+r
\left(M'-1\right)\right)\right.\right.\right.\right.\nn\\
&&\left.\left.\left.\left.-4 r (r-2 M) \Phi ' f_1''(\Phi)\right)-4 r (r-2 M) f_1'(\Phi) \Phi
''\right)\right)\right]\,.\nn
\end{eqnarray}
Equation~\eqref{eqROT} must be solved by imposing regularity
conditions at the center of the star: $\zeta(0)=\zeta_c$,
$\zeta'(0)=0$. We must also require continuity of $\zeta(r)$ at the
stellar radius.  The asymptotic behavior at infinity reads
\begin{equation}
 \zeta\to\zeta_\infty+\frac{2J}{r^3}\,,
\end{equation}
where $J$ denotes the angular momentum. For the solution to be
asymptotically flat, we must impose $\zeta_\infty=0$. This can be
easily achieved by noting that Eq.~\eqref{eqROT} is invariant under
the transformation
\begin{equation}
 \zeta\to\zeta-\eta\,,\quad
 \Omega\to\Omega-\eta\,,
\end{equation}
where $\eta$ is some constant.  Therefore we can proceed as follows
\cite{Hartle:1967he}: (1) integrate Eq.~\eqref{eqROT} imposing
regularity at the center and extract $\zeta_\infty$ and
$\zeta'_\infty$ at some large (but finite) radius $r_{\infty}$; (2)
find the physical value of the angular velocity, i.e.,
$\Omega-\zeta_\infty$. After this translation, $\zeta\to2J/r^3$ at
infinity; (3) compute 
\be
J=-\lim_{r\to\infty}\frac{r^4\zeta'_\infty}{6}\,.
\ee

As we vary $\Omega-\zeta_c$ we obtain models with different specific
angular momentum $J/M^2$. As long as $\Omega\ll \sqrt{M/R_s^3}$, the
slow-rotation approximation is consistent. Ignoring terms of ${\cal
  O}(\Omega^2)$, the moment of inertia is given by $I=J/\Omega$ and it
does not depend on $\zeta_c$, but only on the stellar mass. Therefore
we need to integrate Eq.~\eqref{eqROT} only once in order to obtain
$I$ for a given mass.

With the slowly rotating solution at hand, we can also study the
possibility of ergoregion formation \cite{Schutz:ERGO}. The ergoregion
can be found by computing the surface at which $g_{tt}$ vanishes,
i.e., from Eq.~\eqref{metric_rot}:
\begin{equation}
 -B(r)+\zeta^2 r^2\sin^2\theta=0\,.\label{ergoregion}
\end{equation}
On the equatorial plane we simply have
\begin{equation}
 r\zeta(r)=\sqrt{B(r)}\,,\label{ergo_equatorial}
\end{equation}
and, due to the linearity of the field equations, $\zeta$ will scale
linearly with $\Omega$. Thus, one needs only a single integration in
order to compute the zeros of Eq.~\eqref{ergo_equatorial} as functions
of $\Omega$. For a given value of $\Omega$, there can be no zeros
(i.e. no ergoregion), two distinct zeros (with the ergoregion located
between them) or two coincident zeros. The ``critical frequency'' at
which we have two coincident zeros, say $\Omega_c$, is the minimum
rotation frequency for which an ergoregion exists. The slow-rotation
approximation imposes $\Omega\lesssim\Omega_{\rm ms}$, where the mass
shedding frequency is defined as $\Omega_{\rm ms}\equiv
\sqrt{{M}/{R_s^3}}$, following Hartle's conventions. The existence of
an ergoregion requires $\Omega>\Omega_c$, so an ergoregion can exist
(in the slow-rotation approximation) only if
\begin{equation}
\Omega_c<\Omega_{ms}\,.\label{ergocondition}
\end{equation}

\subsection{Equation of state}

We wish to establish some limits on the parameters of allowed theories
of gravity, given a set of EOS models compatible with our present
knowledge of nuclear physics. Instead of general relativity being
assumed, the parameters characterizing any specific theory will be
constrained based on astrophysical observations. These constraints
will be sensitive to our assumptions on the EOS, that we describe in
this section.

\begin{table}[th!]
\caption{\label{tab:EOS} EOS models used in this work.}
\begin{tabular}{cc}
EOS & Reference \\
\hline \hline 
Polytropic 		& \cite{Damour:1993hw}		\\
FPS 			& \cite{Lorenz:1992zz}		\\
APR			& \cite{Akmal:1998cf}	 	\\
Causal limit	& \cite{Hebeler:2010jx,Koranda:1997}	\\	
\hline
\end{tabular}
\end{table}

A list of the EOS models used in this work is given in
Table~\ref{tab:EOS}.  For code testing purposes, we have considered
the same polytropic model used by Damour and Esposito-Far\'ese
\cite{Damour:1993hw}:
\begin{equation}
\rho=n m_b+K\frac{n_0 m_b}{\Gamma-1}\left(\frac{n}{n_0}\right)^\Gamma \,,\quad P=Kn_0
m_b\left(\frac{n}{n_0}\right)^\Gamma \,,\label{EOSpoly} 
\end{equation}
with $m_b=1.66\times 10^{-24}\,{\rm g}$, $n_0=0.1\,{\rm fm}^{-3}$,
$\Gamma = 2.34$ and $K = 0.0195$.

%
We also considered two nuclear-physics motivated models (FPS
\cite{Lorenz:1992zz} and APR \cite{Akmal:1998cf}, in the standard
nomenclature), which are respectively a soft EOS and a more standard
realistic EOS, as well as for the stiffest possible EOS constructed by
combining the upper limit in the crust-core transition region of
Hebeler et al.~\cite{Hebeler:2010jx} with a causal limit EOS as
in~\cite{Koranda:1997}.
The polytropic model~\eqref{EOSpoly} gives results which are
quantitatively very similar to those for FPS EOS. 

We must remark that the FPS EOS seems to be ruled out by the recent
observation of a neutron star with
$M=(1.97\pm0.04)M_\odot$~\cite{Demorest:2010bx}, at least if we limit
consideration to nonrotating models\footnote{Rapidly rotating neutron
  stars have a larger maximum mass which, using the FPS EOS, is still
  marginally compatible with the observational errors in a small
  region of the parameter space: some rapidly rotating models in Table
  3 of~\cite{Berti:2003nb} have a gravitational mass compatible with
  the value measured in~\cite{Demorest:2010bx}.}  within general
relativity.  However, these observations could be explained in terms
of modified gravity at high density, rather than by invoking a
different EOS. In fact, in some alternative theories the maximum mass
of a neutron star can be sensibly {\em larger} than in general
relativity \cite{Pani:2011mg,Pani:2010vc}. Another important
motivation to use the FPS EOS is to make direct comparison with
previous work. We explicitly checked that our two independent codes
(written in {\sc Mathematica} and {\sc C++}) are in excellent
agreement with Refs.~\cite{Berti:2004ny,1975ApJ...196..653H,Read:2008iy} in the
general relativistic limit\footnote{Incidentally, the moment of
  inertia shown in Fig.~3 of \cite{Yunes:2009ch} is not computed using
  the FPS EOS, as erroneously written in the caption of that
  figure. This explains why our results for the FPS EOS do not agree
  with those in \cite{Yunes:2009ch}.}.

\begin{figure*}[htb]
\begin{center}
\begin{tabular}{cc}
\epsfig{file=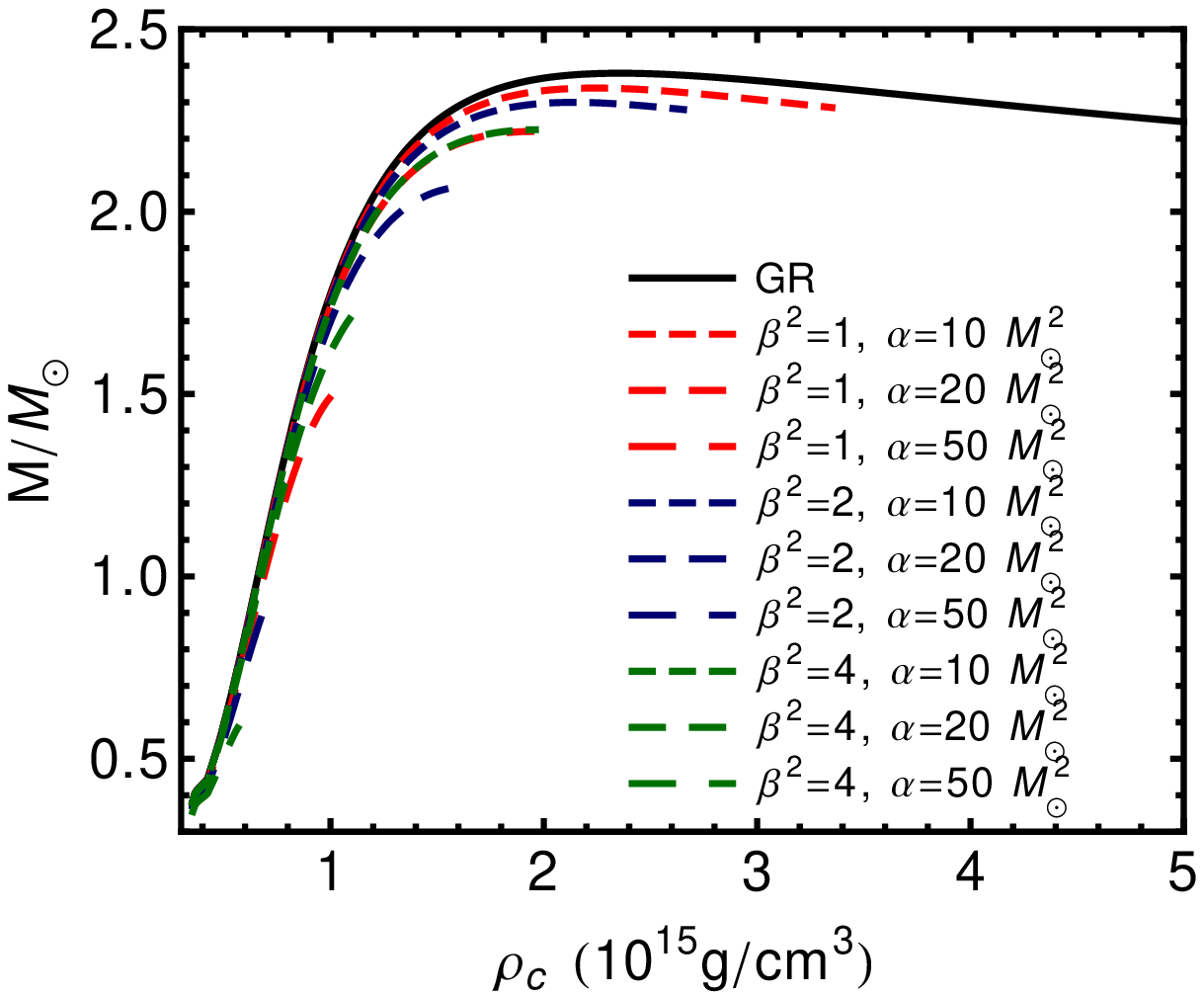,width=7.0cm,angle=0}&
\epsfig{file=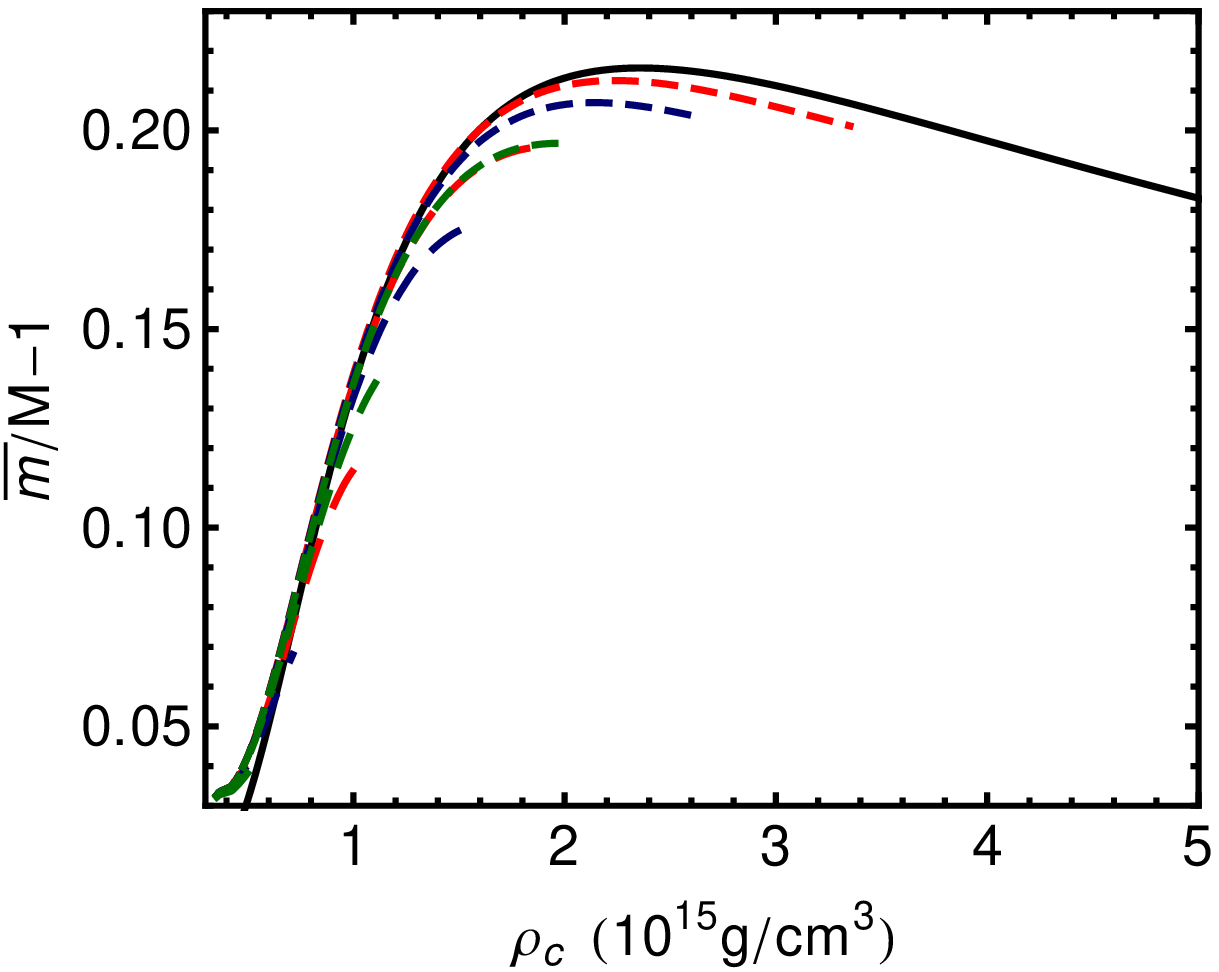,width=7.0cm,angle=0}\\
\epsfig{file=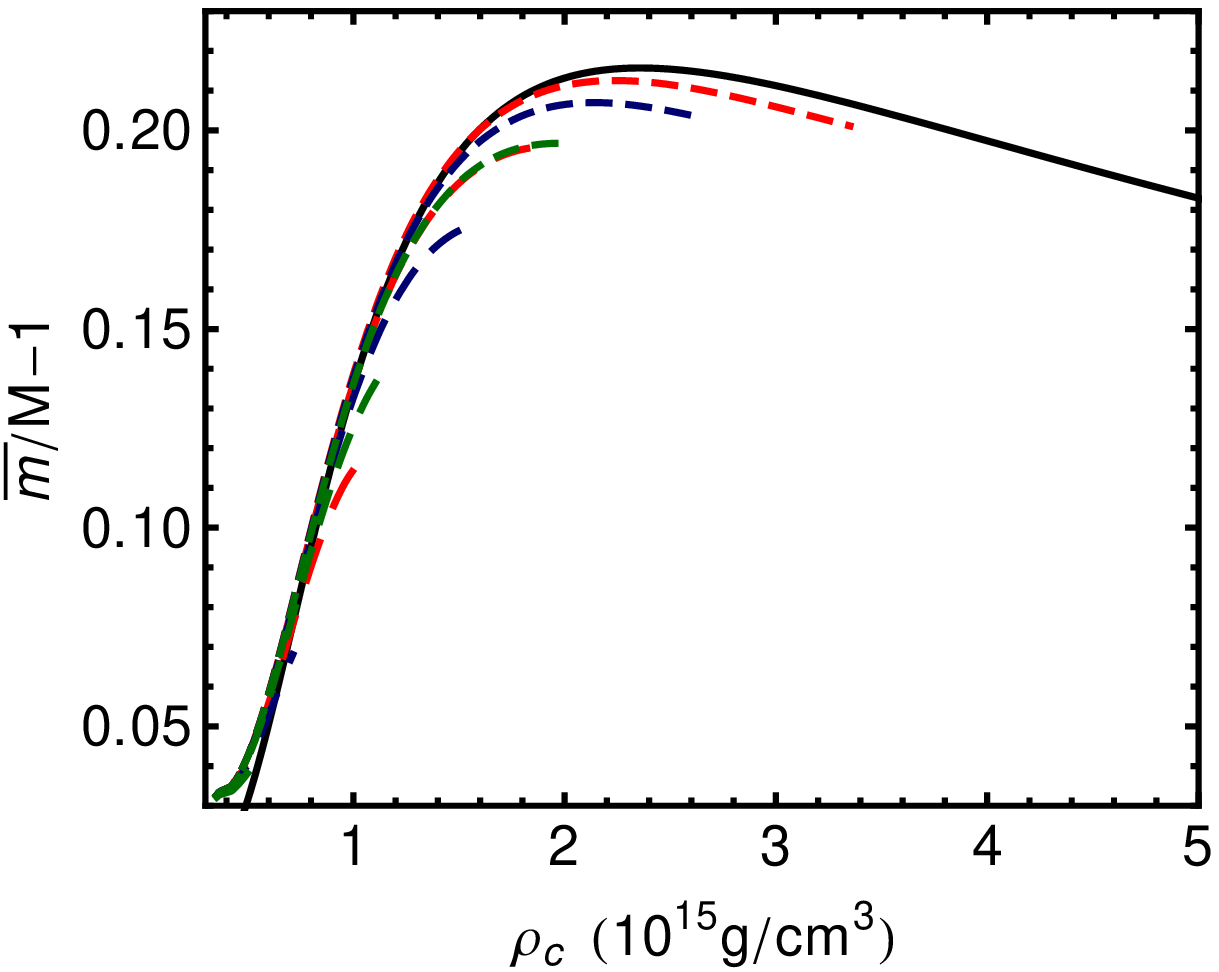,width=7.0cm,angle=0}&
\epsfig{file=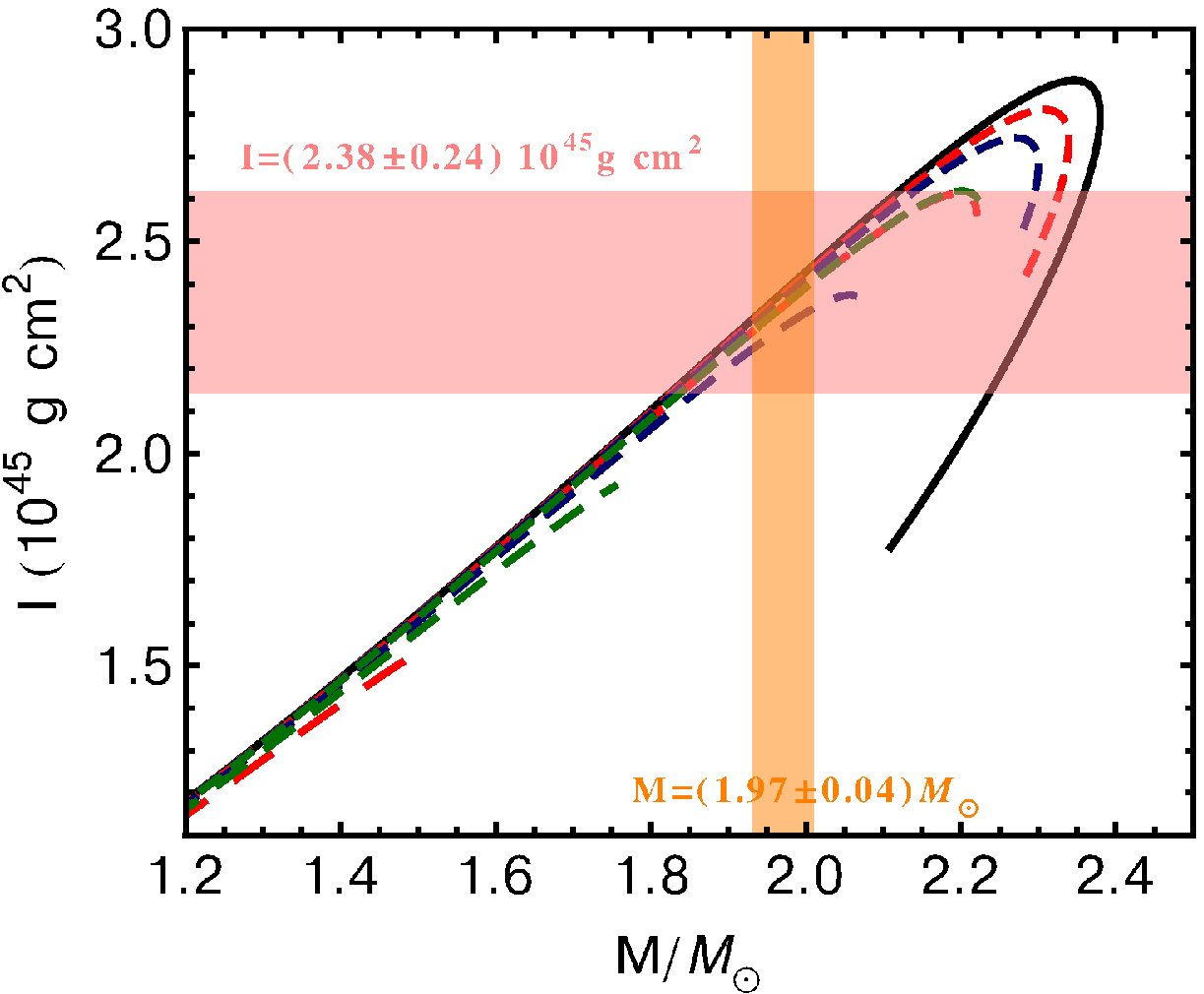,width=7.0cm,angle=0}
\end{tabular}
\caption{Compact star models in EDGB gravity for different values of
  the parameters $\alpha$ and $\beta$, using the APR EOS.  In the
  bottom-right panel we show the recent observation of a neutron star
  with $M\approx 2M_\odot$ and a possible future observation of the
  moment of inertia confirming general relativity within
  $10\%$~\cite{Lattimer:2004nj}. Curves terminate when the
  condition~\eqref{alpha_max} is not fulfilled (cf. also the exclusion
  plot in Fig.~\ref{fig:alpha_max}).
\label{fig:APR}}
\end{center}
\end{figure*}
\begin{figure*}[htb]
\begin{center}
\begin{tabular}{cc}
\epsfig{file=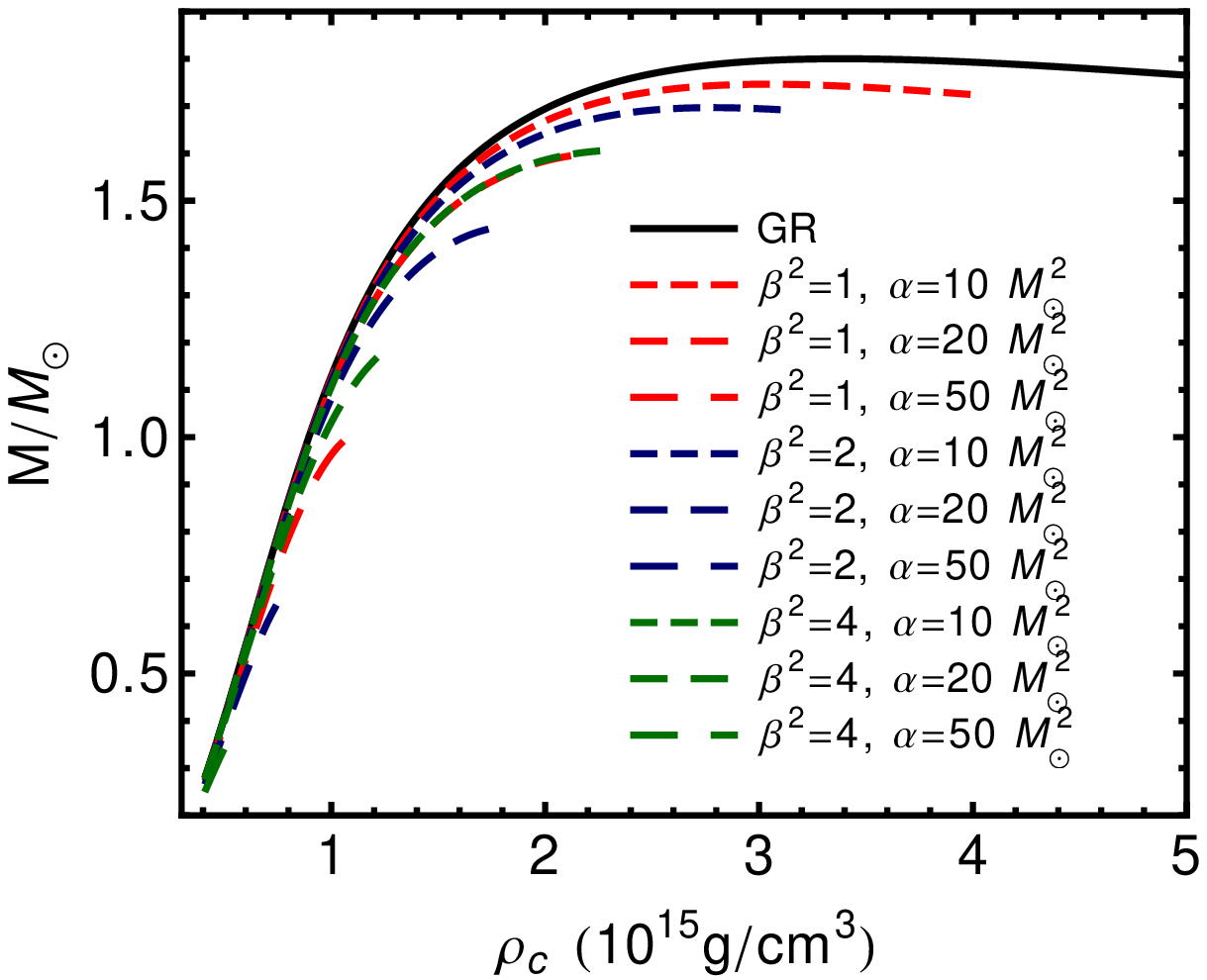,width=7.0cm,angle=0}&
\epsfig{file=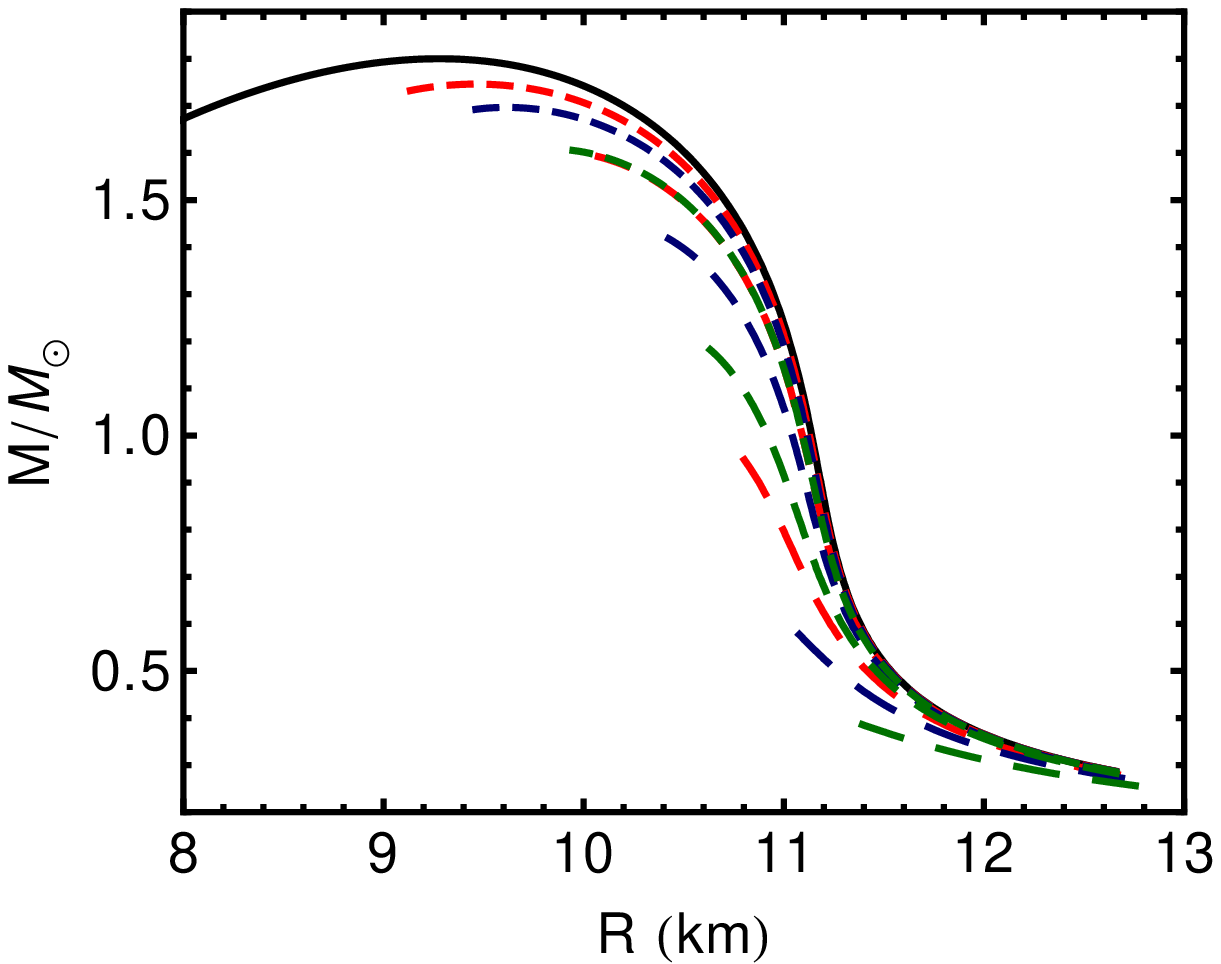,width=7.0cm,angle=0}\\
\epsfig{file=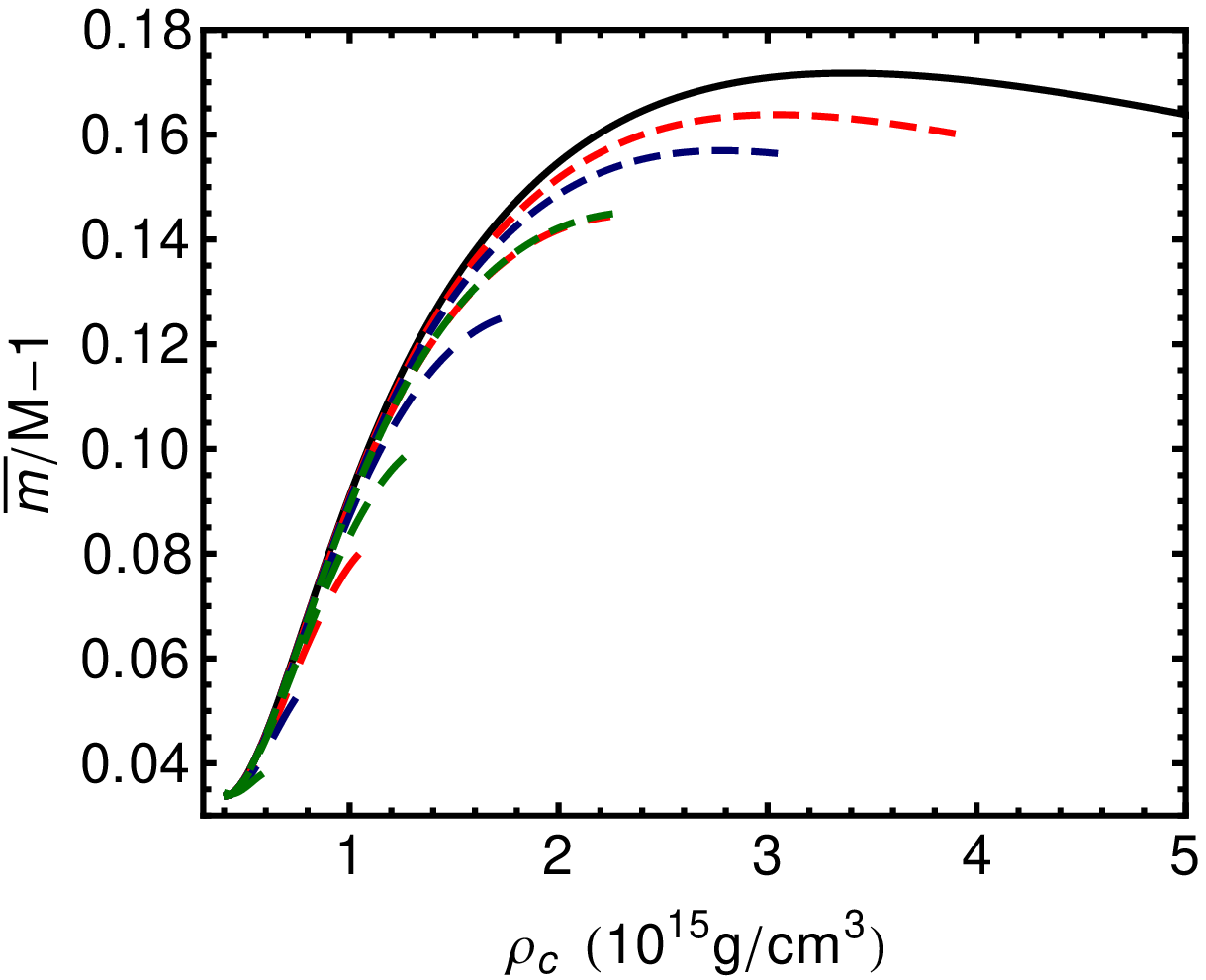,width=7.0cm,angle=0}&
\epsfig{file=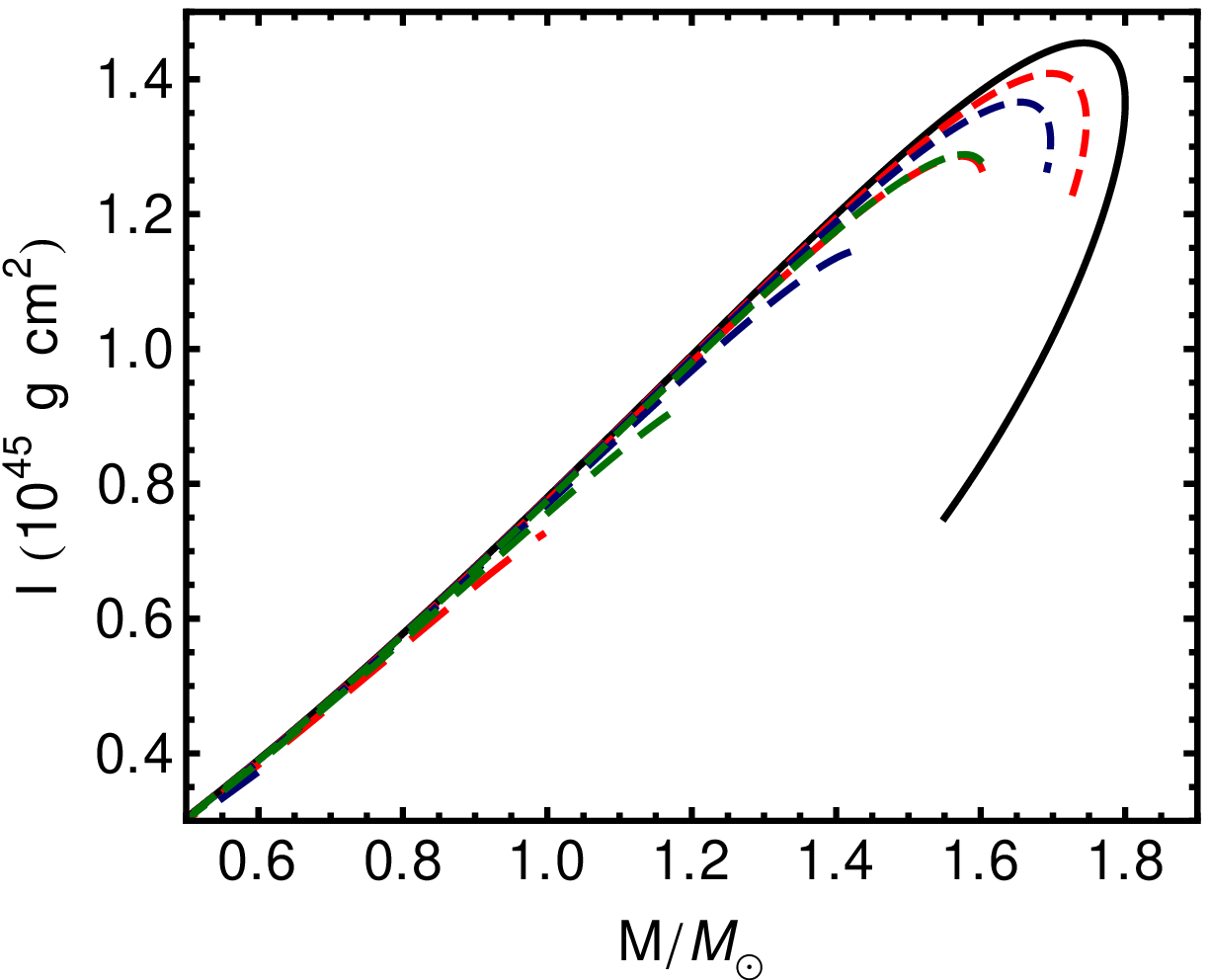,width=7.0cm,angle=0}
\end{tabular}
\caption{Compact star models in EDGB gravity for different values of
  the parameters $\alpha$ and $\beta$, using the FPS EOS. Curves
  terminate when the condition~\eqref{alpha_max} is not fulfilled
  (cf. also the exclusion plot in Fig.~\ref{fig:alpha_max}).
\label{fig:FPS}}
\end{center}
\end{figure*}
 \begin{figure*}[htb]
 \begin{center}
 \begin{tabular}{cc}
 \epsfig{file=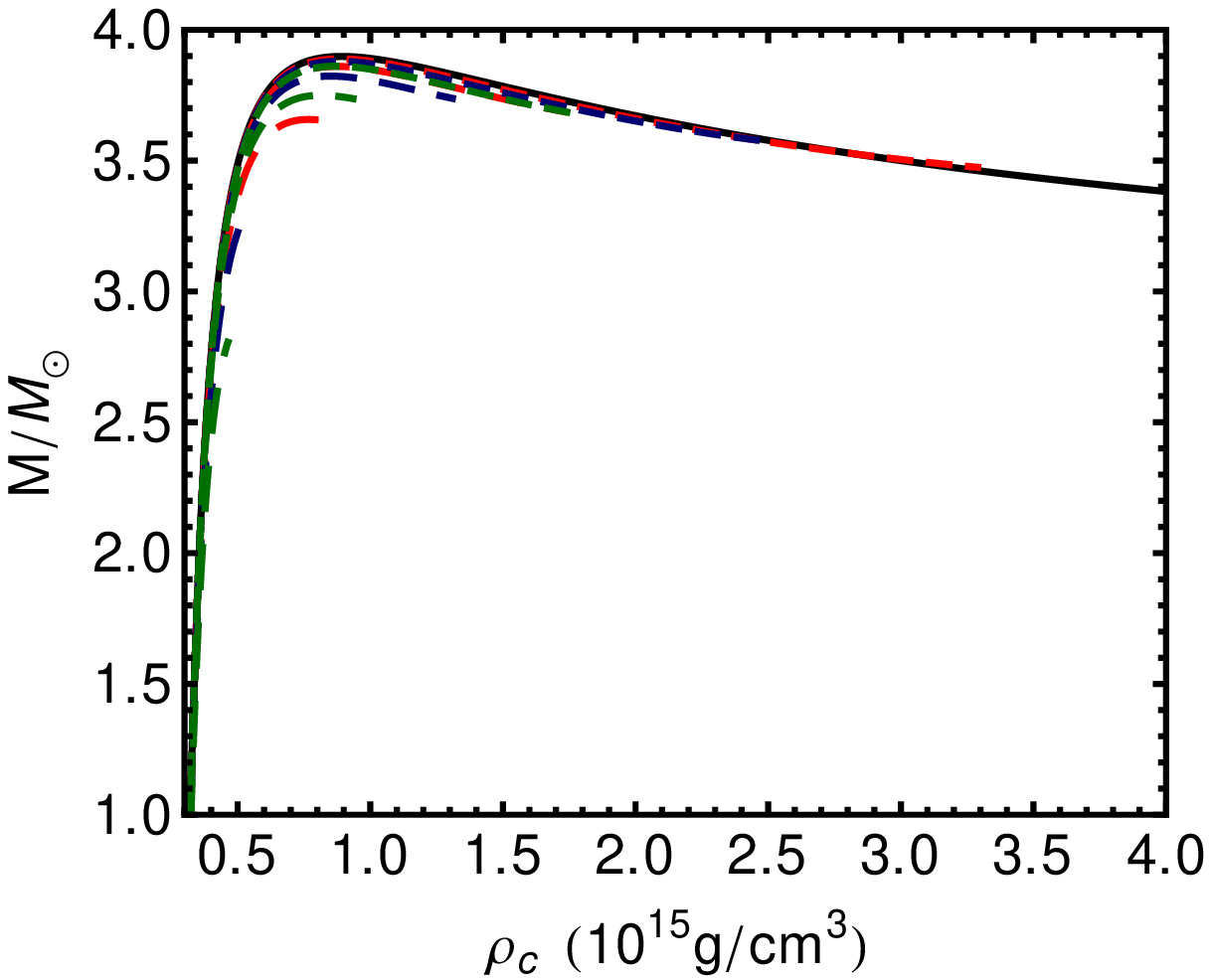,width=7.0cm,angle=0}&
 \epsfig{file=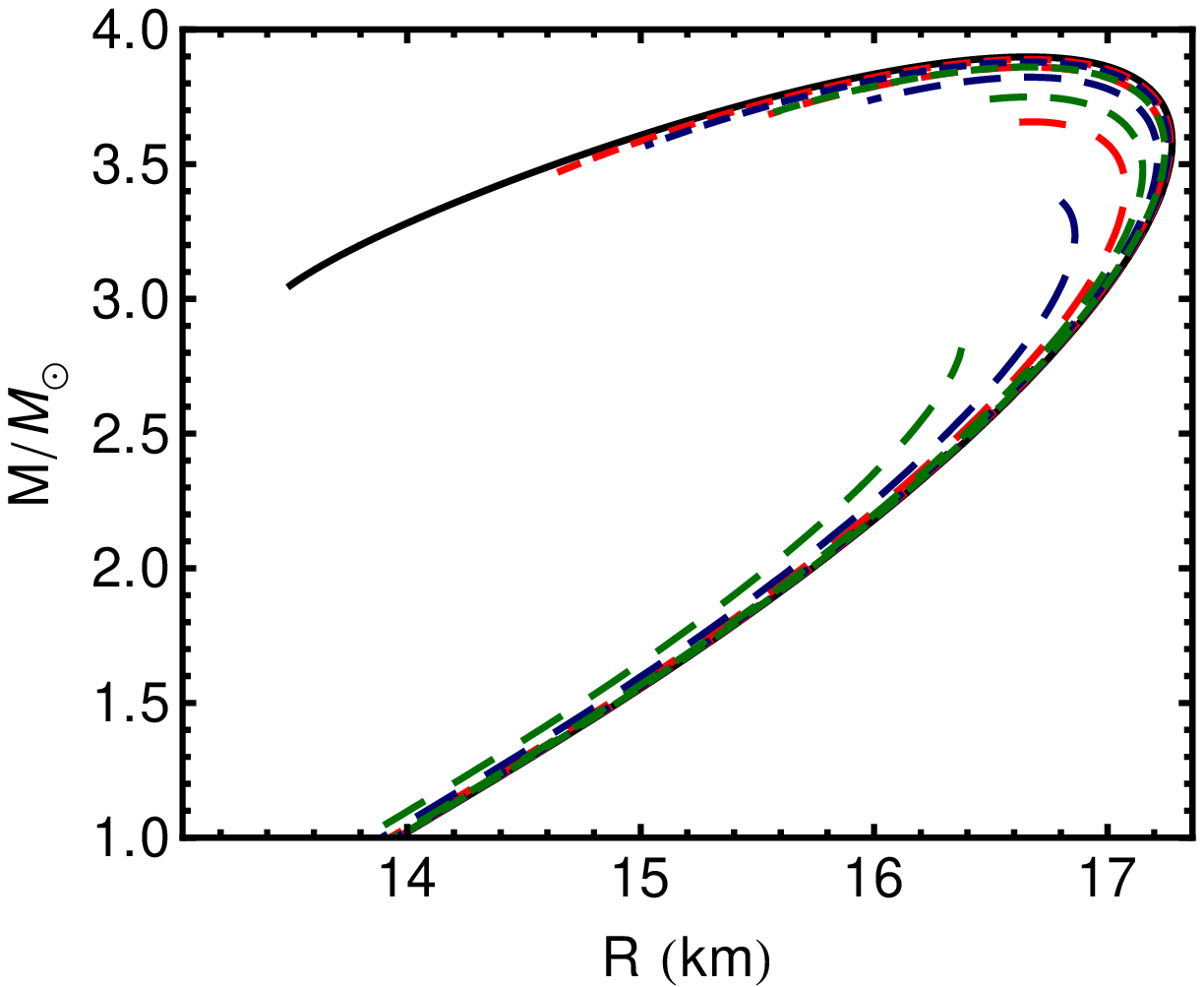,width=7.0cm,angle=0}\\
 \epsfig{file=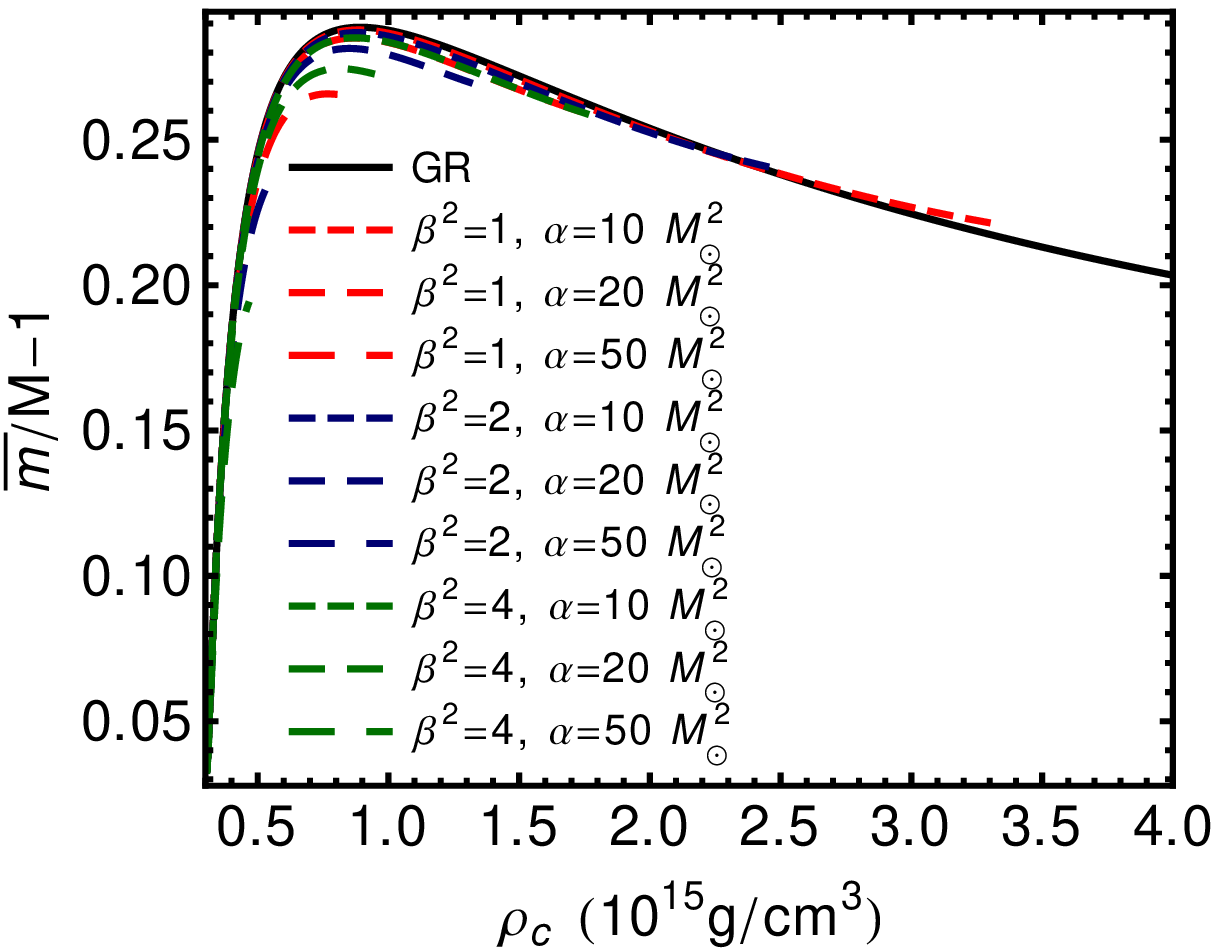,width=7.0cm,angle=0}&
 \epsfig{file=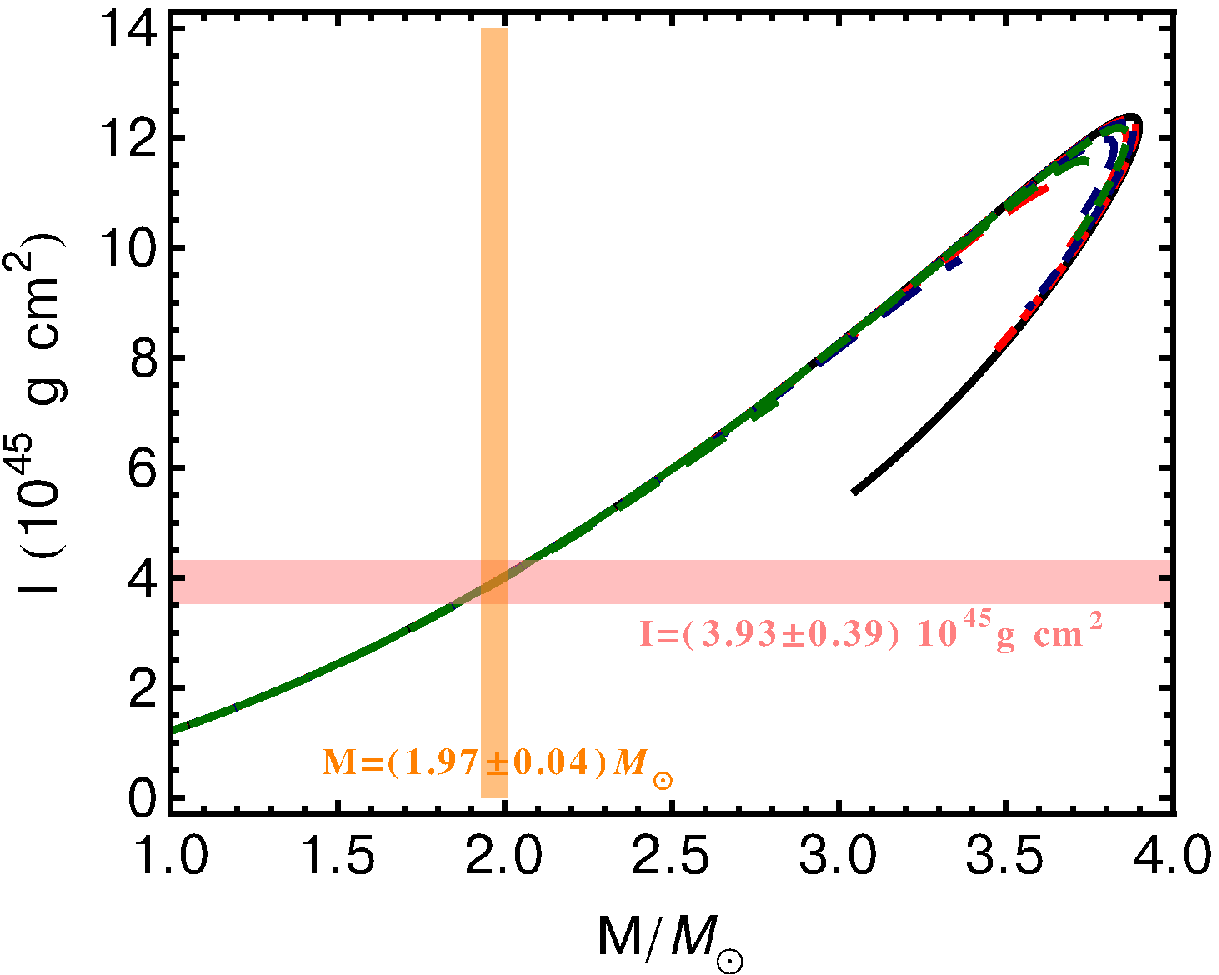,width=7.0cm,angle=0}
 \end{tabular}
 \caption{Compact star models in EDGB gravity for different values of
   the parameters $\alpha$ and $\beta$, using a causal EOS. Curves
   terminate when the condition~\eqref{alpha_max} is not fulfilled
   (cf. also the exclusion plot in
   Fig.~\ref{fig:alpha_max}). \label{fig:Causal}}
 \end{center}
 \end{figure*}
%

\section{Compact stars in Gauss-Bonnet gravity\label{sec:starsEDGB}}

As a first application of the formalism discussed above, in the
remainder of this paper we study neutron stars in EDGB gravity.  We
defer a more general study of the full theory derived from the
Lagrangian~\eqref{lagrangianGB} to future work.

EDGB gravity is obtained from the Lagrangian~\eqref{lagrangianGB} by
considering a real scalar field $\phi=\Phi$ (or $\omega=0$),
$f_0\equiv\kappa=(16\pi)^{-1}$, $V\equiv0$ and
\begin{equation}
f_1\equiv\frac{\alpha}{16\pi} e^{\beta\Phi}\,,\label{f1EDGB}
\end{equation}
where $\alpha$ and $\beta$ are coupling constants. Static black holes
in EDGB gravity only exist when $\alpha$ is positive
\cite{Kanti:1995vq} and for this reason we will only consider
$\alpha>0$. When $\beta=\sqrt{2}$, this theory arises as a low-energy
correction to the tree-level action in heterotic string theory
\cite{Gross:1986mw}. Here we adopt a phenomenological point of view
and consider $\alpha$ and $\beta$ as free (real) parameters. We will
show by an explicit calculation that, under reasonable assumptions for
the nuclear EOS, the observation of compact stars with certain
observed properties (such as mass, radius or moment of inertia) leads
to the existence of rather stringent exclusion regions in the
two-dimensional $(\alpha\,,\beta)$ parameter space.

Some results are shown in Figs.~\ref{fig:APR}-\ref{fig:Causal} for
different EOS models and different values of $\alpha$ and $\beta$.  In
each figure, the top two panels show the mass-density relation and the
mass-radius relation for static (nonrotating) stars. The bottom-left
panel shows the binding energy as a function of the central
density. In the bottom-right panel we display the moment of inertia as
a function of (gravitational) mass.

In our numerical calculations, we observed that the scalar field in
the interior of the star is always small: in special cases it can be
as large as $\Phi\sim10^{-2}$, but more typically
$\Phi\lesssim10^{-4}$ in most of the parameter space.  In the
small-field limit, $\Phi\ll1$, the coupling $f_1$ in
Eq.~\eqref{f1EDGB} can be Taylor-expanded:
\begin{equation}
 {16\pi}f_1(\Phi)\sim{\alpha}+{\alpha\beta}\Phi\,.\label{f1series}
\end{equation}
Since the first term is constant and the GB term is a topological
invariant, the first nonvanishing corrections arise from the second
term. Therefore, in the small-field limit the equilibrium structure
depends only on the product $\alpha\beta$ of the coupling
constants. This is confirmed by our numerical results in
Figs.~\ref{fig:APR}-\ref{fig:Causal}: for instance the lines
corresponding to $\alpha=20M_\odot^2$, $\beta^2=1$ and $\alpha=10
M_\odot^2$, $\beta^2=4$ both correspond to the same
$\alpha\beta=20M_\odot^2$, and indeed they lie almost exactly on top
of each other.
A similar degeneracy will occur for any other functional form of
$f_1(\Phi)$, provided that the scalar field remains small everywhere,
so that a Taylor expansion similar to Eq.~\eqref{f1series} holds. 
In this sense, most of our results remain valid for a generic function
$f_1(\Phi)$, and not only for EDGB gravity.

\begin{figure}[htb]
\begin{center}
\begin{tabular}{c}
\epsfig{file=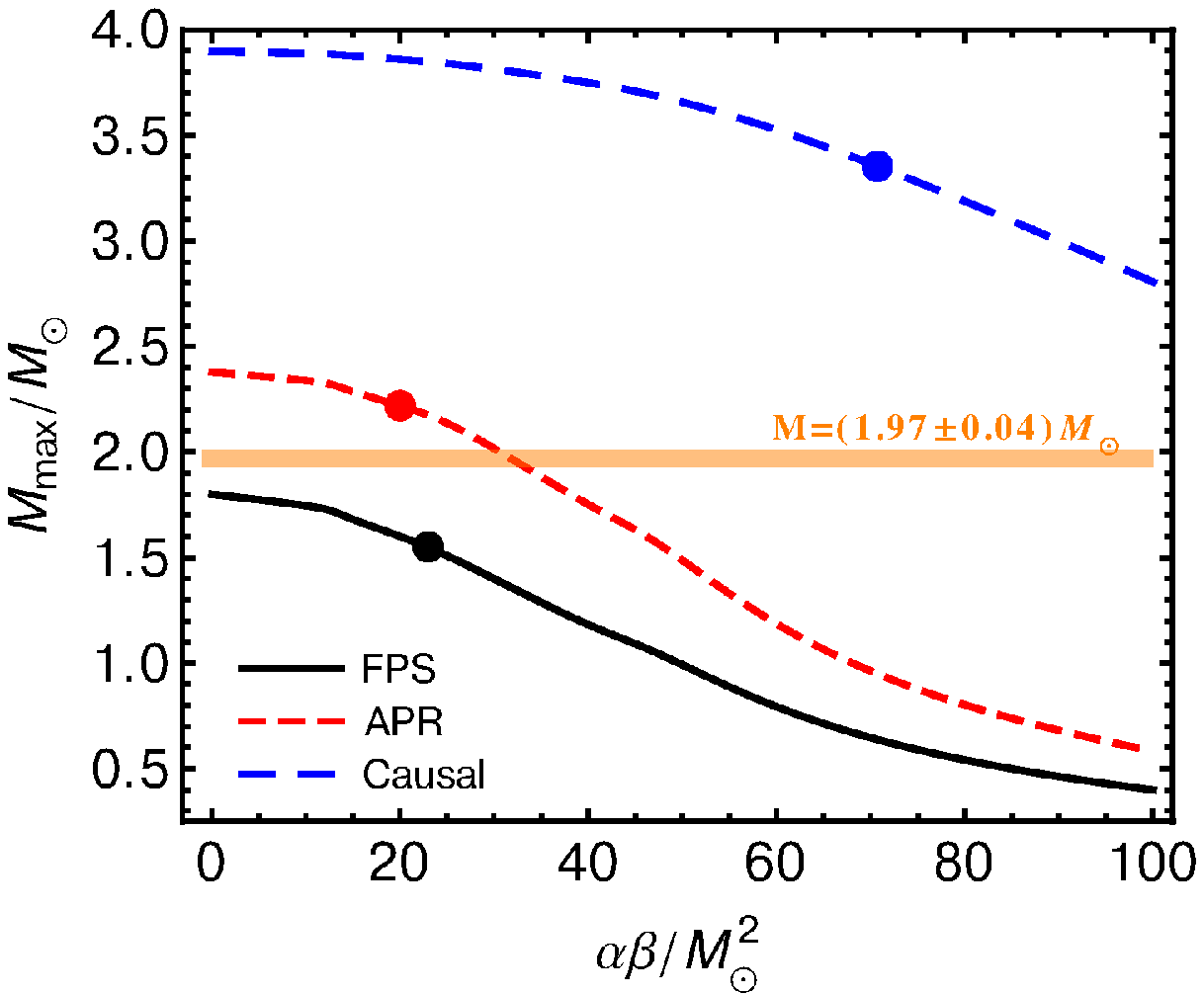,width=7.0cm,angle=0}
\end{tabular}
\caption{Maximum mass as a function of the product $\alpha\beta$ of
  the EDGB coupling parameters, for different EOS models and in the
  nonrotating case (cf. the main text for details). To the left of the
  filled circle, this maximum mass corresponds to the radial stability
  criterion; to the right, it corresponds to the maximum central
  density for which we can construct static equilibrium models. The
  recent measurement of a neutron star with $M\approx2
  M_\odot$~\cite{Demorest:2010bx} is marked by a horizontal line.
  Only the combination $\alpha\beta$ is bounded, due to the
  approximation $\Phi\ll1$: cf. Eq.~\eqref{f1series}.
\label{fig:M_max}}
\end{center}
\end{figure}

It is clear from Figs.~\ref{fig:APR}-\ref{fig:Causal} that, regardless
of the EOS, for $\alpha>0$ the coupling to the dilaton tends to {\em
  reduce} the importance of relativistic effects. Indeed, as shown in
Fig.~\ref{fig:M_max}, the maximum gravitational mass $M_{\rm max}$
monotonically decreases as a function of the product $\alpha\beta$ of
the EDGB coupling parameters. Thus in EDGB gravity (as well as in
general relativity) soft EOS models, like FPS, should be ruled out by
observations of high-mass neutron stars.  This is similar to what
happens in gravitational-aether theory~\cite{Kamiab:2011am} and in
Einstein-aether theory~\cite{Eling:2007xh,Eling:2007xh}.

For small values of the product $\alpha\beta$, the maximum mass in
Fig.~\ref{fig:M_max} corresponds to a local maximum in the
mass-density relation (cf.~the upper left panels of
Figs.~\ref{fig:APR}-\ref{fig:Causal}).  In general relativity these
local maxima (or, equivalently, inversion points in the mass-radius
diagram) correspond to marginally stable equilibrium configurations,
and solutions to the right of the first maximum are unstable to radial
perturbations (see e.g.~\cite{Shapiro:1983du}). We conjecture that the
same property should hold also for extended scalar-tensor
theories. This was proved for particular self-gravitating
configurations involving scalar fields
\cite{Gleiser:1988ih,Gleiser:1988rq,Harada:1997mr}, but a more
detailed stability analysis would be desirable (see also the
discussion in \cite{Horbatsch:2010hj}).  In EDGB theory, spherically
symmetric solutions can be constructed only up to a maximum central
density $\rho_c^\text{max}(\alpha,\beta)$, for reasons explained
below: see in particular the discussion around
Eq.~\eqref{alpha_max}. For large $\alpha\beta$ this maximum central
density is such that the first local maximum in the mass-density curve
is never reached. In Fig.~\ref{fig:M_max}, all values to the left of
the solid circles correspond to a maximum mass obtained from the
radial stability criterion. Values to the right of the solid circles
correspond instead to the mass $M_{\rm max}$ obtained at the critical
value of $\rho_c$ beyond which we cannot find spherically symmetric
perfect fluid solutions anymore.

%

\subsection{Constraints on the EDGB couplings}
In the near future, observations of the double pulsar may provide
measurements of the moment of inertia to an accuracy of $\sim 10\%$
\cite{Lattimer:2004nj} (but see Ref.~\cite{Iorio:2007yz} for some criticism).
Furthermore, precise observations of the
mass-radius relation could be obtained from thermonuclear X-ray burst
\cite{Guver:2011qw,Guver:2011js}.
These observations could be used in the context of a Bayesian
model-selection framework to place strong constraints on EDGB gravity
and, more generally, to remove the degeneracy between different EOS
models and different proposed modifications of general relativity.

Nevertheless, even without assuming any particular EOS, we can set
rather stringent theoretical constraints on the EDGB parameters.
Indeed, as shown in Figs.~\ref{fig:APR}-\ref{fig:Causal}, depending on
$\alpha$ and $\beta$, there is a maximum value of the central density
$\rho_c$, above which no compact star models can be constructed. For a
given central density, the critical value of $\alpha\beta$ can be
computed analytically in the small $\Phi$ limit, as follows.  We first
compute the series expansion~\eqref{series_center} up to ${\cal
  O}(r^2)$. The resulting expressions are not very illuminating, but
in general the series coefficients contain square roots, whose
argument must be positive to ensure the existence of physical
(real-valued) solutions. When $\Phi_c\ll1$, by imposing this ``reality
condition'' we find
\begin{eqnarray}
 &\alpha^2\beta^2<&\frac{1}{7776 \pi  P_c^4 \rho_c}
\left[128 \rho_c^3-27 P_c^2 \rho_c+288 P_c \rho_c^2+54 P_c^3\right.\nn\\
&&\left.-2\sqrt{ (3 P_c+\rho_c)
\left(3 P_c-8 \rho_c\right)^2 (3 P_c+4 \rho_c)^3}\right].\label{alpha_max}
\end{eqnarray}
For a given value of $\alpha\beta$, the condition above implies that a
maximum central density, $\rho_c^\text{max}$,
exists. 

\begin{figure}[htb]
\begin{center}
\begin{tabular}{c}
\epsfig{file=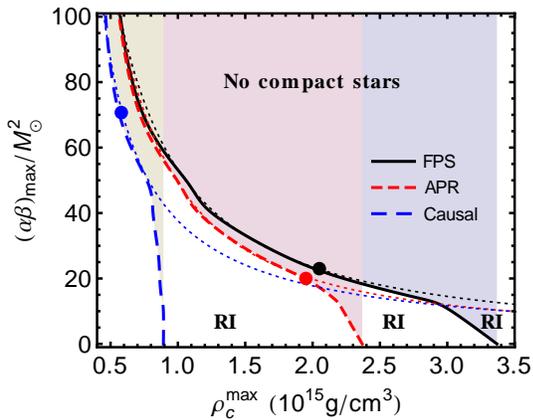,width=7.0cm,angle=0}
\end{tabular}
\caption{Exclusion plot in the $(\rho_c^{\rm max},\,\alpha\beta)$
  plane for nonrotating models. No compact star solutions can be
  constructed in the region above the dotted lines
  (cf. Eq.~\eqref{alpha_max}). In the region above the thick lines
  (labeled by ``RI'', for ``radial instability''), static
  configurations should be unstable against radial perturbations: see
  the main text for details. Filled circles indicate the maximum
  central density of radially stable stars in general relativity.
\label{fig:alpha_max}}
\end{center}
\end{figure}

Equation~\eqref{alpha_max} is in good agreement with numerical
results, as shown in Fig.~\ref{fig:alpha_max}.  This figure is
basically an exclusion plot: it shows the maximum allowed values of
$\alpha\beta$ as a function of the maximum central density
$\rho_c^\text{max}$ for different EOS models and nonrotating stars.
For small values of $\alpha\beta$ (i.e., on the right of the figure),
a local maximum in the mass-density relation is reached and the
maximum central density $\rho_c^\text{max}$ corresponds to the local
maximum of $M(\rho_c)$, i.e. to the first inversion point in the
mass-radius relation. If our stability conjecture is correct, no
stable static configurations can be constructed in the region
\emph{above} these lines. The local maximum is never reached to the
\emph{left} of the points marked by filled circles. In this case, the
$M(\rho_c)$ curves terminate before reaching a local maximum, and the
maximum central density $\rho_c^\text{max}$ is simply the point where
the equilbirum sequence terminates.  Dotted lines correspond to the
analytical prediction~\eqref{alpha_max}, which agrees very well with
the numerical value at which we cannot compute equilibrium models
anymore. The bottom line is that no static models (either stable or
unstable) can be constructed in the shadowed regions above these
exclusion lines.

The quadratic EDGB corrections are expected to
be stronger in high-density (high-curvature) regions, so the most
stringent bounds should come from the stiffest EOS models. Indeed,
among the models we consider, the strongest and weakest constraints
come from the Causal EOS and from the FPS EOS, respectively. 
%


The bounds on the central density can be translated into constraints
on the maximum mass, which is an observable quantity. As shown in
Fig.~\ref{fig:M_max}, for a given EOS the maximum mass is a
monotonically decreasing function of $\alpha\beta$.

The requirement that the maximum mass $M_{\rm max}$ supported by the
theory should be larger than some trusted observed value, can place
a direct upper bound on $\alpha\beta$. In Table~\ref{tab:constraints},
we consider $M_{\rm max}\gtrsim1.4 M_\odot$, $M_{\rm max}\gtrsim1.7
M_\odot$ and $M_{\rm max}\gtrsim1.93 M_\odot$ (which is the lower
bound of the recent observation in \cite{Demorest:2010bx}) and we
translate them into upper bounds on $\alpha\beta$ using the data
plotted in Fig.~\ref{fig:M_max}. The Causal EOS models predict an
unrealistically large maximum mass, so all neutron stars with $M_{\rm
  max}\lesssim 2.8 M_\odot$ would place very mild constraints on
alternative theories, and we omit this EOS from
Table~\ref{tab:constraints}.

\begin{table}[th!]
\caption{\label{tab:constraints} Constraints on the EDGB parameters
  from an observation of a nonrotating neutron star with mass $M$. For
  the values of $M$ we consider, constraints using the Causal EOS
  would allow values of $\alpha\beta$ larger than $100 M_\odot^2$.}
\begin{tabular}{cccc}
EOS & $M_{\rm max}\gtrsim1.4M_\odot$ & $M_{\rm max}\gtrsim1.7M_\odot$ & $M_{\rm max}\gtrsim1.93M_\odot$ \\
\hline \hline 
FPS 			& $\alpha\beta\lesssim 30.1M_\odot^2$ & $\alpha\beta\lesssim 13.9M_\odot^2$ & no models		\\
APR			& $\alpha\beta\lesssim 50.3M_\odot^2$ & $\alpha\beta\lesssim 41.9M_\odot^2$ & $\alpha\beta\lesssim 33.6M_\odot^2$	 	\\
\hline
\end{tabular}
\end{table}
For the ``standard'' value of the EDGB coupling ($\beta=\sqrt{2}$), if
we consider the APR EOS as our ``best candidate'' EOS for a
(nonexotic) neutron star interior within general relativity, the
results in Table~\ref{tab:constraints} imply $\alpha\lesssim 23.8
M_\odot^2$.

This bound should be compared to the bound on $\alpha$ that comes from
requiring the existence of black hole solutions in the theory
\cite{Kanti:1995vq,Pani:2009wy}. For $\beta=\sqrt{2}$, this
requirement implies
\begin{equation}
\frac{\alpha}{M_\odot^2}\lesssim70\left[\frac{M_{BH}}{10M_\odot}\right]^2\,,
\label{EDGBBH}
\end{equation}
where $M_{BH}$ is the black hole mass. The observation of black holes
with $M_{BH}\approx8 M_\odot$ (such as Cyg X1) constrains
$\alpha\lesssim44 M_\odot^2$.  The constraints on $\alpha$ coming from
observations of compact stars (cf. Table~\ref{tab:constraints}) are
already smaller by a factor $\sim 2$ than those coming from the
existence of stellar black holes, and they could become even more
stringent in the near future.

\begin{figure}[htb]
\begin{center}
\begin{tabular}{c}
\epsfig{file=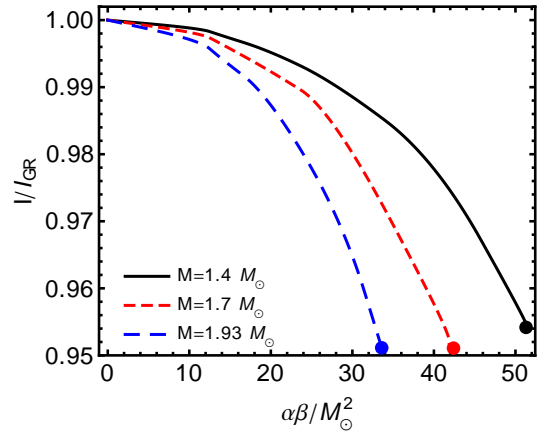,width=7cm,angle=0}
\end{tabular}
\caption{Moment of inertia normalized by its value in general
  relativity for the APR EOS at fixed values of the gravitational
  mass. Curves terminate at the bounds listed in
  Table~\ref{tab:constraints} (corresponding to the filled
  circles). The deviations from general relativity are always smaller
  than $5\%$. \label{fig:I_constraints}}
\end{center}
\end{figure}

We may hope that future observations of the moment of inertia could
place even tighter bounds on the theory. Unfortunately this seems
unlikely. To understand why, we can either look at the bottom right
panel of Fig.~\ref{fig:APR} or at Fig.~\ref{fig:I_constraints}, where
we show the moment of inertia for the APR EOS (normalized by its value
in general relativity) as a function of $\alpha\beta$ for fixed values
of the stellar mass. As it turns out, for values of $\alpha\beta$
smaller than those listed in Table~\ref{tab:constraints}, the moment
of inertia can deviate from the general relativistic value by $5\%$ at
most. The precision of future observations is expected to be $\sim
10\%$ in optimistic scenarios~\cite{Lattimer:2004nj}.  Therefore, at
least for EDGB gravity, the most stringent constraints on
$\alpha\beta$ should come from mass measurements, rather than from
measurements of the moment of inertia.
%

Let us mention, for completeness, that we also studied the possibility
of the formation of an ergoregion for slowly rotating stars in EDGB
gravity. For our `realistic'' EOS models the
condition~\eqref{ergocondition} is never met, and therefore no
ergoregion can form outside the star. This is because, for
phenomenologically viable parameters, the relativistic effects in this
particular theory are actually {\em smaller} than in general
relativity. We can anticipate that other sectors of the general theory
described by the Lagrangian~\eqref{lagrangianGB} could enhance
relativistic effects and favor the existence of the ergoregion, with
important implication for the stability of these solutions
\cite{Schutz:ERGO}. A more detailed analysis will be presented
elsewhere.

\section{Conclusions and outlook}\label{sec:conclusions}
Neutron stars are very promising laboratories to constrain
strong-curvature corrections to general relativity. New proposed
theories of gravity are usually tested against weak field observations
and cosmological data, or by studying the existence and nature of
black hole solutions. Our main goal in this paper was to develop a
formalism for a comprehensive study of stellar structure in a broad
class of alternatives to Einstein's general relativity. We focused on
a class of theories (``extended scalar-tensor theories'') where
quadratic curvature corrections, nonminimal couplings and
parity-violating terms are coupled to standard gravity through a
single scalar field. Particular cases of this model include, but are
not limited to, quadratic gravity, EDGB gravity, generic scalar-tensor
theories and $f(R)$ theories (via their correspondence with
scalar-tensor theories). We wrote down the field equations for static
and spherically symmetric perfect-fluid stars in the general case, as
well as the leading-order corrections in a slow-rotation
expansion. For a given model and a given central density, the
formalism allows us to obtain the mass, radius, binding energy and
moment of inertia of compact stars. In future work we will show how
these theoretical predictions can be compared to observations in order
to constrain the parameter space of ``extended scalar-tensor'' and
other alternative theories.

As a first application of the formalism we studied stellar structure
in EDGB gravity. We found that, in general, the GB coupling tends to
{\it reduce} relativistic effects in compact stars. We also showed
that there is an exclusion region in the two-dimensional plane of the
GB coupling parameters beyond which no compact star solutions can be
constructed: cf. Eq.~\eqref{alpha_max}, Fig.~\ref{fig:M_max} and
Fig.~\ref{fig:alpha_max}.

Stability requirements for static models and future observational data
could constrain the theory even further.  The existence of high-mass
neutron starss put the most stringent constraints on EDGB gravity
(cf. Table~\ref{tab:constraints}).  As it turns out, these bound are
tighter (by a factor of a few) than the bound coming from the
existence of black hole solutions in EDGB theory, given in
Eq.~\eqref{EDGBBH}. They are also tighter than the bounds that could
come from future precision measurements of the moment of inertia.

In this sense, to our knowledge, the existence of large-mass neutron
stars provides the best constraint on the EDGB coupling parameters
obtained so far. Further explorations of stellar structure and better
observational data on the mass-radius relation (see
e.g.~\cite{Guver:2011qw,Guver:2011js}) have the potential to exclude a
larger region of the parameter space of alternative theories. There is
of course the possibility that theoretical and observational work may
give us hints on how to modify general relativity to make it
compatible with the standard model, which would be even more exciting.

\noindent
{\bf \em Acknowledgments.}
We thank Nico Yunes for useful discussions.  This work was supported
by the {\it DyBHo--256667} ERC Starting Grant, by NSF Grant
PHY-0900735, by NSF CAREER Grant PHY-1055103, and by FCT - Portugal
through PTDC projects FIS/098025/2008, FIS/098032/2008,
CTE-AST/098034/2008.

\bibliography{stars_alternative_theories}
\end{document}